\documentclass[aps,groupedaddress,showpacs]{revtex4}

\usepackage{graphicx}
\usepackage{epsfig}

\def\wbar{\bar{w}}
\def\begeq{\begin{equation}}
\def\endeq{\end{equation}}

\def\begeqar{\begin{eqnarray}}
\def\endeqar{\end{eqnarray}}

\def\wbar{\bar{w}}

\def\bz{{\bar z}}
\def\bpsi{{\bar \psi}}
\def\bphi{{\bar\phi}}

\begin{document}
\bibliographystyle{apsrev}


\title{
Scattering amplitudes in non-Fermi liquid systems}


\author{A.~Koutouza$^1$, F. Lesage$^2$ and H.~Saleur$^{1,3,4}$}
\affiliation{${}^1$Department of Physics,
University of Southern California, Los Angeles, CA 90089-0484\\
$^2$ Centre de recherches math\'ematiques, Universit\'e de Montr\'eal,
C.P. 6128, succ. centre-ville, Montr\'eal, P.Q., Canada, H3C-3J7 \\
${}^3$Physikalisches Institut, Albert-Ludwigs-Universit\"at,
 D-79104 Freiburg, Germany\\
${}^4$ Service de Physique Th\'eorique,
CEN Saclay, Gif Sur Yvette F91191}


\date{\today}

\begin{abstract}
By a mix of form-factors and analyticity techniques, we determine
some fundamental scattering amplitudes in non-Fermi liquid systems.
These include the reflection and transmission amplitudes for Laughlin
quasiparticles at a point contact between two $g={1\over 2}$
Luttinger liquids, and the $e\rightarrow e$, $lqp\rightarrow lqp$ and
$e\rightarrow e_{edge}$ at a point contact between a Fermi liquid and a
$g={1\over 3}$ Luttinger liquid (all liquids chiral or not)
. These results are obtained in closed
form, and give rise to rather simple expressions for the
probabilities of the most basic processes of non Fermi liquid physics
at these special values of the couplings. 
\end{abstract}

\maketitle


\section{Introduction}

The purpose  of this paper is to address the calculation
of  scattering amplitudes in systems exhibiting non-Fermi liquid 
physics. 
In such systems, scattering processes involving electrons
(more precisely, the quasiparticles adiabatically connected with the 
bare electrons in the Landau Fermi liquid theory, with charge $e$ and
spin $1/2$) 
can give rise to the appearance of other kinds of quasiparticles. A
characteristic example of this phenomenon (to be discussed in details 
below) occurs when  electrons are  injected
from a Fermi liquid into a Luttinger liquid, and can give rise to
various combinations of Laughlin quasiparticles and quasiholes \cite{SCW}.
Another example occurs in  the  spin $1/2$ 
two channel Kondo problem: at the low energy fixed point, 
the amplitude for an electron to be scattered into an electron (or any finite combination of electrons and holes) is actually zero, hence giving rise to `
unitarity puzzles', which are solved by recognizing that the scattering takes place entirely into the spinors sector \cite{Kondoref}.  

The determination of these amplitudes is conceptually a very
important question, as one may argue they are the essence of non Fermi liquids physics. Technically however, it is a very difficult
one, which has been tackled mostly in perturbation theory around
the high energy fixed point \cite{FS}. The 
purpose of this paper is to present 
some non perturbative attempts at calculating these amplitudes in  a case
where the problem is integrable - and especially simple.  

Our initial attempts were  based on  the technique of form-factors \cite{FFref}, 
and this paper  builds on
earlier works in this area \cite{Giuseppe,LSS,LS}.   As we will see, the answer 
to the physical questions of interest involves the determination of
correlation functions in a massless field theory with a boundary
interaction. 
In the paper \cite{LSS} it 
was shown that such correlation functions could be determined with
remarkable accuracy in the case of operators with no anomalous
dimension, like the current and the stress energy tensor. In the
paper \cite{LS}, preliminary attempts were made to determine similar 
correlators in the case of vertex operators with non trivial,
anomalous dimensions. From a technical point of view, the present
work is a follow up of that paper, and will use similar
regularization techniques. It will turn out however that for the
specific questions asked - for instance, what is the probability that
an electron from a (chiral) Fermi liquid 
 tunnels into a  (chiral) 
Luttinger $\nu={1\over 3}$ liquid as an edge electron, the form-factor results do not, in
practice, give reliable results for all scales of energy. We will
thus have to make considerable use of another non peturbative
(``analytic'')
approach  pioneered by Chatterjee and Zamolodchikov \cite{CZ} to obtain what
will turn out to be results in closed form.   

Two main applications are presented: the problem of an impurity in a one
dimensional Luttinger liquid, and the problem of a point contact between a Fermi liquid and
a Luttinger liquid.
These two problems are 
related  with the boundary sine-Gordon model, with  which we assume the reader to be familiar. Mostly for technical reasons, we restrict in this paper 
to the case $g={1\over 2}$. Although this is a ``free fermion theory'', the amplitudes of interest involve operators which are not local in terms of the fermions, and whose correlators are already quite complex. The physical features are not expected to depend strongly on $g$, so we expect our results to shed light on the general situation; moreover, $g={1\over 2}$ is  of direct relevance to the case of a contact between a Fermi liquid and
 a $\nu={1\over 3}$ edge \cite{CF}.  

In the case of the impurity in a Luttinger liquid, or, 
equivalently, tunneling between two identical chiral Luttinger liquids (section 2), 
we discuss both the reflection and transmission amplitude for a 
single Laughlin quasiparticle. In the case of tunneling between a
 chiral Luttinger liquid and a Fermi liquid (section 3), we discuss the
 $e\rightarrow e$, the $lqp\rightarrow lpq$ and the $e\rightarrow
 e_{edge}$ amplitudes. Here $e_{edge}$ denotes the edge electron, which can also be considered as the bound state of three Laughlin quasiparticles.

 Exact expressions are obtained in all cases. For the ease of the
 reader, we gather these expressions here. In the first situation we 
 have
 \begin{eqnarray}
     G_{T}(p,p_{B})&\propto& p^{-1/2}\left(1+{ip\over
     p_{B}}\right)^{-1/2}~~~~~~~~~~~~~~\hbox{Transmission of lqp}\nonumber\\
     G_{R}(p,p_{B})&=&0, ~ ~ p_{B}\neq 0~~~~~~~~~~~~~~~~~~~~~~~~~~~~
     \hbox{Reflection of lqp} 
   \end{eqnarray}
 while in the second 
 \begin{eqnarray}
     G(p/p_{B})&=&F\left({1\over 2},{1\over 2},1;-i{p\over p_B}\right)~~~~~~~~~~~~~~~~~~~~~~~~~~~~~~~~~~
         \hbox{$1e\rightarrow 1e$ }\nonumber\\
	 G(p,p_{B})&\propto &p^{-2/3}
	    F\left({1\over 2},{1\over 2},{1\over 3};-i{p\over p_B}\right)~~~~~~~~~~~~~~~~~~~~~\hbox{$1lqp\rightarrow 1lqp$ 
	     }\nonumber\\
	 G(p,p_{B})&\propto& p^{1/4}
\left[{\Gamma(1/2)\Gamma(3/4)\over \Gamma(5/4)} F\left({1\over 2},{1\over 2},{5\over 4},-i{p\over p_B}\right)+{\Gamma(1/2)\Gamma(-1/4)\over\Gamma(1/4)}
F\left({1\over 2},{1\over 2},{1\over 4};-i{p\over p_B}\right)\right]
	     ~~~~~~~~\hbox{$1e\rightarrow 1e_{edge}$}
	     \end{eqnarray}
Some technical issues are relegated to the appendices.

To conclude this introduction, we would like to point out that the mere questions we want to answer have to be defined very carefully, a task we do not completely  undertake in this paper. Indeed, in the massless theories we are discussing, there are ambiguities in the set of scattering states one uses as a ``physical basis''. For the one dimensional Fermi liquid, the choice of electrons and holes is canonical, but for general Luttinger liquids, this is not so. For instance, for the $g={1\over 3}$ case, a choice based on charges ${1\over 3}$ (Laughlin quasihole) and $-1$ (edge electron) 
quantas has been studied in great details \cite{Kareljan}. Other choices, eg using charges ${1\over 3}$ and $-{1\over 3}$ (Laughlin quasiparticle) quantas, should be possible as well: none appears more fundamental than the other, but the choice has to be carefully specified whenever one discusses, for instance, unitarity. Whatever the choice, the vertex operators (in the bosonized formalism) one uses as ``creation/annihilation'' operators of these quantas are not one particle operators in the usual sense. They do not simply add or subtract a single quasi-particle from a many particle state, but have a more complicated action due to their non trivial commutators. This can make the mere definition of a scattering amplitude quite non trivial \cite{Kareljan}.

\section{Impurity within a Luttinger liquid}

\subsection{Generalities}

\begin{figure}
\vspace{0.3cm}
\begin{center}
\epsfig{file=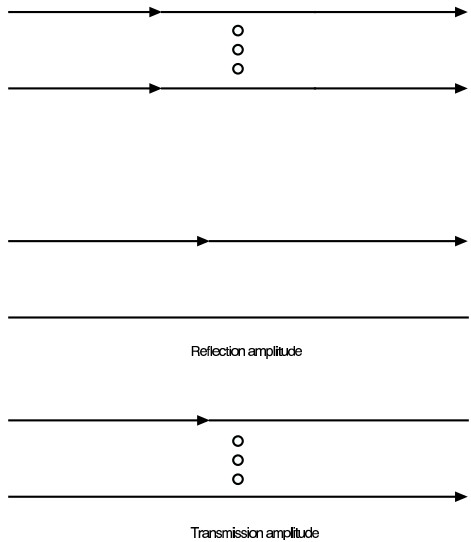,scale=0.4}  
\caption{\label{fig1} Tunneling between two identical chiral Luttinger
liquids. What we call reflection corresponds to a lqp remaining
within the same liquid, and what we call transmission amplitude
corresponds to a lqp jumping from one liquid to the other. }
\end{center}
\end{figure}

We consider a problem with two identical chiral Luttinger liquids
and tunneling between them induced by a gate voltage - the set up is 
represented on the figure \ref{fig1}.

The Lagrangian reads
\begin{equation}
L={1\over 4\pi} \partial_x\phi_1(\partial_t-\partial_x)\phi_1+{1\over 
4\pi} \partial_x\phi_2(\partial_t-\partial_x)\phi_2
+v\delta(x)\cos[\sqrt{\nu}(\phi_1-\phi_2)],
\end{equation}
where the fields $\phi_{1},\phi_{2}$ are associated with each of the 
two Luttinger liquids. By forming linear combinations 
\begin{equation}
    \phi'={\phi_{1}+\phi_{2}\over\sqrt{2}},~~~
    \phi={\phi_{1}-\phi_{2}\over\sqrt{2}}
\end{equation}
the field $\phi'$ decouples, while the  dynamics of the  $\phi$ field is
determined by the Lagrangian
\begin{equation}
L={1\over 4\pi} \partial_x\phi(\partial_t-\partial_x)\phi
+v\delta(x)\cos(\sqrt{2\nu}\phi).
\end{equation}
The coupling constant $v$ describes the strenght of the tunneling process.
The normalization is such that one has the
propagator ($T$ the time ordering operator)
\begin{equation}
\langle T\phi(x,t)\phi(0,0)\rangle=-{1\over
4\pi}\ln(x-t+i\epsilon sign(t)) .
\end{equation} 
It is then convenient to fold this problem onto the half line
$x\in[-\infty,0]$, with action (here $\Phi=\phi(x,t)+\phi(-x,t)$ is a non chiral boson)
\begin{equation}
    A={1\over 8\pi} \int_{-\infty}^{0} dx
    \left[(\partial_x\Phi)^2)-(\partial_{t}\Phi)^2 \right]+v
    \cos[\sqrt{\nu/ 2}\Phi(0)].
    \end{equation}
The boundary interaction has physical dimension $d=\nu$. In the
following we consider mostly the case $\nu={1\over 2}$. 

The two amplitudes we wish to consider are what we call the {\sl  reflection}
and
{\sl transmission} amplitudes for Laughlin quasiparticles. The creation
operator of a Laughlin quasiparticle in the first Luttinger liquid
is proportional to $e^{-i\sqrt{\nu}\phi_1}$ so the reflection
amplitude  
(corresponding to a Laughlin quasiparticle going through within the
same chiral Luttinger liquid) is  
\begin{eqnarray}
 G_{R}&=&\langle T~
e^{i\sqrt{\nu}
\phi_{1}(x_{2}>0,t_{2})}e^{-i\sqrt{\nu}\phi_{1}(x_{1}<0,t_{1})}\rangle\nonumber\\
&=& {1\over (x_2-x_{1}-t_{2}+t_{1}+i\epsilon~ sign(t_{2}-t_{1}))^{\nu/2}}
\langle T~
e^{i\sqrt{\nu/2}
\phi(x_{2}>0,t_{2})}e^{-i\sqrt{\nu/2}\phi(x_{1}<0,t_{1})}\rangle
\end{eqnarray}
while the transmission amplitude (corresponding to a Laughlin 
quasiparticle jumping from one  
 chiral Luttinger liquid to the other ) reads
\begin{eqnarray}
 G_{T}&=&\langle T~
e^{i\sqrt{\nu}
\phi_{2}(x_{2}>0,t_{2})}e^{-i\sqrt{\nu}\phi_{1}(x_{1}<0,t_{1})}\rangle \nonumber\\
&=& {1\over (x_2-x_{1}-t_{2}+t_{1}+i\epsilon ~sign(t_{2}-t_{1}))^{\nu/2}}
\langle T~
e^{-i\sqrt{\nu/2}
\phi(x_{2}>0,t_{2})}e^{-i\sqrt{\nu/2}\phi(x_{1}<0,t_{1})}\rangle .
\end{eqnarray}
The main technical difference between these two amplitudes is the fact that
exponentials of the field $\phi$  have opposite signs in the
reflection case and identical signs in the transmission case. 
Restricting to $\nu={1\over 2}$, the two objects we have to determine
are thus
\begin{equation}
g_{R,T}(1,2)=\langle T~
\exp(\pm i
\phi(x_{2}>0,t_{2})/2)\exp( -i\phi(x_{1}<0,t_{1})/2)\rangle .
\end{equation}
To proceed, we first use the mapping to the folded
theory on the half line. A field at $x<0$ is an `in', right moving
field, while a field at $x>0$ now becomes an `out', left moving field.
The new correlators are then given by
\begin{equation}
g_{R,T}(1,2)=\langle T~
 \exp(\pm i
\phi_{L}(-x_{2},t_{2})/2)\exp (-i\phi_{R}(x_{1},t_{1})/2)\rangle .
\end{equation}

The general approach to compute such correlators is the massless form-factors
technique \cite{Giuseppe,LSS}. Let us therefore  recall some basic facts about integrable 
boundary field theories.  The hamiltonian is diagonalized using a 
quasiparticle basis, 
with left and right massless particles $e=\pm p= e^\beta$, carrying a 
label $\pm$ corresponding to soliton or antisoliton. The bulk 
scattering 
of these quasiparticles is $S=-1$. Their boundary scattering is
\begin{equation}
R^+_-=R^-_+={e^{\beta-\beta_B}\over 
e^{\beta-\beta_B}+i},~~~R_+^+=R_-^-={i\over e^{\beta-\beta_B}+i}
\end{equation}
where $e^{\beta_B}\propto v^2$. 

The form-factors of the vertex operators  can in principle be
obtained by taking the massless limit  of the formulas in 
\cite{BernardLeclair}. Subtleties arise in taking this limit however.
Let us first recall the massive result. Restricting to $|\alpha|
<1$, the basic formula in \cite{BernardLeclair} reads, setting
$U_{\alpha}=\exp(i\alpha\Phi)$, 
\begin{eqnarray}
{\langle 0|U_\alpha(x,t)|\theta_{2n}\ldots\theta_1\rangle^{+\ldots+,-\ldots 
-}\over  \langle 0|U_\alpha|0\rangle}&=&
(-)^{n(n-1)/2}\left({\sin\pi\alpha\over i}\right)^n\nonumber\\
&\times&\exp\left(-im\sum_{k=1}^{2n}(t\cosh\theta_{k}-x\sinh\theta_{k})\right)
\exp\left(\alpha \sum_{k=1}^n 
(\theta_{n+k}-\theta_k)\right)\nonumber\\
&\times&{\prod_{1\leq k<j\leq n} \sinh{\theta_k-\theta_j\over 
2}\sinh{\theta_{n+k}-\theta_{n+j}\over 2}\over \prod_{1\leq k,j\leq n} 
\cosh{\theta_{n+k}-\theta_j\over 2}}\label{formfactori}
\end{eqnarray}
The vacuum expectation value itself is obtained using techniques of
\cite{Lukyanovetal} and is of the form
$<0|U_\alpha|0>=(am)^{\alpha^{2}}C(\alpha)$ where $a$ is a UV 
cut-off and $C(\alpha)$ a numerical constant whose expression is
exactly known. 
Notice that the form factors are neutral, that is they are non zero
only acting on states whose total solitonic charge vanishes exactly. 

The massless limit is obtained by letting rapidities go to infinity 
setting $\theta=\pm\theta_{0}\pm\beta$ where
$\theta_{0}\rightarrow\infty$, and letting at the same time the mass 
go to zero with $me^{\theta_{0}}\rightarrow 2$ (the choice of this
constant is non universal, and can be absorbed in a global shift of
the rapidities). For example, let us take the massless limit of the form-factor by
choosing $p$ of the $n$ particles with charge $+$ to be R movers, and
$q$ to be $L$ movers (with $p+q=n$) and similarly, $u$ particles
among the $n$ particles with charge $-$ to be R movers, $v$ particles
to be L movers (with $u+v=n$). As $\theta_{0}\rightarrow\infty$, the 
leading behaviour of the ratio (\ref{formfactori})  behaves as
$$
\exp[\alpha\theta_{0}(p-q-u+v)]
{e^{\theta_{0}pq}e^{\theta_{0}uv}\over
e^{\theta_{0}pv}e^{\theta_{0}uq}}
$$
This can be rewritten as $e^{\theta_{0}[\alpha^{2}-(\alpha+q-v)^{2}]}$.
Taking the prefactor $\langle 0|U_\alpha|0\rangle$ into account, it 
thus follows that the form-factors all vanish in the massless limit. 
The calculation of  correlators is thus bound  to be  a complex task: what happens is that, while the   form-factors vanish in that
limit,  
the integrals over 
rapidities all diverge at low energy  due to the
presence of infinitely many soft modes, an ``IR catastrophe'',
making the contributions to correlators of undertermined form
 $0\times \infty$. The proper way to 
proceed would be  to keep track of this divergence carefully, by making  
 calculations at finite 
mass, and then considering the limit
when the length scales get much smaller than the inverse of this 
mass. It is not clear however how many form-factors would have to be 
taken into account in order to get accurate results - probably an
infinity - and this procedure clearly is not efficient.

The procedure we will  use below bypasses this difficulty,
and consists in working 
directly within the massless theory, essentially by considering  
ratios of correlators, which gets rid of the overall normalization 
problem.
It turns out that the ratios (\ref{formfactori}) have different
behaviours depending on $\alpha$ and the balance of charges. 
In particular, the limit of the ratio is always finite when the charges are
individually balanced in the left and right sector. In that case, the
ratio factorizes, and one can individually define L and R versions as
(setting $V_{\alpha}^{R,resp. L}=e^{i\alpha\phi_{R,L}}$)
\begin{eqnarray}
{\langle 0|V_\alpha^{R,~resp.~L}(x,t)|\beta_{2n}\ldots\beta_1\rangle_{R,~resp.~L}^{+\ldots+,-\ldots 
-}\over  \langle 0|V_\alpha|0\rangle}&=&
(-)^{n(n-1)/2}\left({\sin\pi\alpha\over i}\right)^n\nonumber\\
&\times&\exp\left(-i(t\mp 
x)\sum_{k=1}^{2n}e^{\beta_k}\right)\exp\left(\pm\alpha \sum_{k=1}^n 
(\beta_{n+k}-\beta_k)\right)\nonumber\\
&\times&{\prod_{1\leq k<j\leq n} \sinh{\beta_k-\beta_j\over 
2}\sinh{\beta_{n+k}-\beta_{n+j}\over 2}\over \prod_{1\leq k,j\leq n} 
\cosh{\beta_{n+k}-\beta_j\over 2}}\label{formfactor}
\end{eqnarray}
For $|\alpha|<{1\over 2}$, this is in fact the only case where the
ratio has a non zero limit. 
Our procedure thus  seems  well defined  when
$|\alpha|< {1\over 2}$. When ${1\over 2}<|\alpha|<1$, some of the 
ratios (\ref{formfactori}) diverge in the massless limit, and handling
out the form-factors series  is thus probably
even more difficult. We have not tackled this case, but see the
appendix for analytical techniques. 

The case $|\alpha|={1\over 2}$ is special. In addition to the case
where L and R sectors are separately neutral,  the ratio
(\ref{formfactori}) also has a finite limit when the balance of
charges is $+1$ in the R sector and $-1$ in the L sector. For
instance
\begin{equation}
    {\langle 0|V_{1/2}^{L}
V_{1/2}^{R}|\beta_{2},\beta_{1}\rangle_{R,L}^{+,-}\over
\langle 0|V_{1/2}|0\rangle}=-{2\over i}
\exp\left(-i(t-x)e^{\beta_2}-i(t+x)e^{\beta_{1}}\right)
\end{equation}
while the same expression for $V_{-1/2}$ vanishes in the massless
limit.

Finally, the case $|\alpha|=1$ is also  special, even in the massive case. 
This is because $\sin\alpha\pi$ vanishes, while the constant
$C(\alpha)$ diverges. The best way to understand what happens 
is to make connection with fields with $\alpha=1$ and fermion
operators in a free fermion theory. In the massless limit, one finds 
that $V_{\pm 1}^{R,L}\propto \psi_{R,L}(resp.~\psi^{\dagger}_{R,L}$). 
This is discussed in more details below.

The normalization of multiparticle states in (\ref{formfactor}) is  
$<\beta|\beta'>=2\pi\delta(\beta-\beta')$. We denote below the 
creation operators $Z_{L,R}^{\dagger a}(\beta)$ such that eg 
$Z^{\dagger a}_{L,R}(\beta)|0>=|\beta>^a_{R,L}$. Left and right
creation/annihilation
operators anticommute. 

The boundary interaction can be conveniently handled by moving to an 
Euclidian description, such that the impurity interaction becomes a
boundary interaction, the boundary lying along the imaginary time
axis. One then makes a a Wick rotation, and quantizes the theory
along the boundary axis instead. The coordinate along the boundary,
$y$, is related with the time by $y=it$. The whole effect of the boundary
interaction is then taken into account by the  
 boundary state \cite{GZ}, whose expression in the integrable basis
 is extremely simple:
\begin{equation}
|B>=\exp\left[\int_{-\infty}^\infty {d\beta\over 2\pi} 
K^{ab}(\beta_B-\beta)Z^{\dagger a}_L(\beta)Z^{\dagger 
b}_R(\beta)\right]
\end{equation}
Here, the $K$ matrix is simply related with the $R$ matrix and given 
by
\begin{equation}
K^{++}=K^{--}={ie^{\beta-\beta_B}\over 
e^{\beta-\beta_B}+1},~~~K^{+-}=K^{-+}={-i\over e^{\beta-\beta_B}+1}
\end{equation}

\subsection{The transmission amplitude}

A calculation involving up to two particles gives then the following 
expression, where we neglected some trivial normalization factors
(note that  the L particles are only annihilated
by the $L$ field and the $R$ ones by the $R$ field)
\begin{eqnarray}
g_{T}(1,2)&=&\left<\exp(-i\phi_{L}(-x_2,y_{2})/2)\exp(-i\phi_R(x_1,y_1)/2)\right>
\nonumber\\
&\propto& \left\{1+\int {d\beta\over 2\pi} 
\exp\left[e^{\beta}(x_1-x_2+i(y_1-y_2))\right]
{2\over 1+e^{\hat{\beta}}}\right.
\nonumber\\
&+&
\int {d\beta_1\over 2\pi} {d\beta_2\over2\pi}
{\exp\left[(e^{\beta_1}+e^{\beta_2})(x_1-x_2+i(y_1-y_2))\right]
\over \cosh^2 {\beta_1-\beta_2\over 2}}\times\nonumber\\
&\times &\left. \left[ {1\over 1+e^{\hat{\beta}_1}}{1\over 
1+e^{\hat{\beta}_2}}e^{\hat{\beta}_2-\hat{\beta}_1}
-
{e^{\hat{\beta}_1}\over 
1+e^{\hat{\beta}_1}}{e^{\hat{\beta}_2}\over 
1+e^{\hat{\beta}_2}}
\right]+\ldots\right\}\label{formal}
\end{eqnarray}
where $\hat{\beta}=\beta-\beta_B$, and the dots denote processes 
involving three and more particles in the expansion of the boundary 
state. As mentioned brielfy earlier, this integral is divergent at low energies 
$\beta_i\rightarrow -\infty$ due to the proliferation of low energy
particles, and the expression
as it stands cannot be used. Integrals converge at high energy since 
$x_{2}>0,x_{1}<0$. 

To proceed, we use a technique \cite{FFref} experimented in this context  in \cite{LS}.
Denote $p_B=e^{\beta_B}$ and observe  that the low energy  limit 
is also the limit $v\rightarrow\infty (p_B\rightarrow \infty)$, where the field $\phi$
sees Dirichlet boundary conditions, and thus we do know that 
the correlator should go as 
$$
g_{T}(1,2)_{p_B=\infty}\propto {1\over (x_2-x_1+i(y_2-y_1))^{1/4}}
$$
corresponding to a totally diagonal scattering. The trick to obtain 
finite results will thus be to consider not $g_T$ itself but the ratio 
$g_T(1,2)/g_T(1,2)_{p_B=\infty}$. Writing to first non trivial order
$g_T(1,2)=1+I_1+\ldots$, we thus consider the new approximation to the 
correlator
\begin{equation}
g_{T}(1,2)\approx  {1\over (x_2-x_1+i(y_2-y_1))^{1/4}} 
{1+I_1(p_B)+\ldots\over 1+I_1(\infty)+\ldots}
\end{equation}
and expand the ratio to first order (ie, two $\beta$ integrals)
\begin{equation}
g_{T}(1,2)=  
{1\over (x_2-x_1+i(y_2-y_1))^{1/4}} 
\left[1+I_1(p_B)-I_1(\infty)+\ldots\right]
\end{equation}
This procedure happens to  give finite integrals, as is easily seen 
at this order, since the subtraction of $I_1(\infty)$
cancels the divergence of the first integral in
(\ref{formal}). It follows that, to leading order, the full 
correlator of interest reads, after the Wick rotation
\begin{equation}
G_{T}(1,2)={1\over (x_2-x_1+i(y_2-y_1))^{1/2}} \left\{1-
{1\over\pi} \int_{0}^{\infty} {du\over
p_{B}+u}e^{-u(x_2-x_1+i(y_2-y_1))}+\ldots\right\}
\label{expan}
\end{equation}
After a few simple manipulations using the Laplace representation of 
the square root prefactor, we end up with 
\begin{equation}
G_{T}(p,p_B)=\frac{1}{\sqrt{\pi p}}\left[
1-{1\over\pi}\int_{0}^{p/p_{B}}{du\over 1+u}\left(1-{p_{B}\over
p}u\right)^{-1/2}+\ldots\right]
\end{equation}
In the low energy (that is, low energy of incident particles)
limit $p/p_{B}\rightarrow 0$, corresponding to  
total transmission, $G_{T}\propto
p^{-1/2}(1+\hbox{cst} \frac{p}{p_{B}})$, as expected. In the high energy limit
$p/p_{B}\rightarrow\infty$,  $G_{T}\propto p^{1/2}(1+\hbox{cst}\ln(p_B/p))$. 
This latter result clearly does not make much sense, as one expects
$G_{T}$ to vanish in this limit.

A higher order calculation is perfectly possible, and would extend
the degree of validity of the form-factors result towards high
energies, while giving an essentially exact result at low energies.
For any finite order however, the form-factors result seem to always 
become unreliable at sufficiently high energy, and behave like the
first order in that limit. 
Rather than pursuing this direction (which might be the only one
avaialble for other values of $g$), it is better  to observe that the quantity $G_{T}(1,2)$ can be evaluated 
in closed form. This follows from a chain or arguments relating it
to a very
elegant calculation of the one point function of the spin operator in
the Ising model with a boundary magnetic field carried out by
Chatterjee and 
Zamolodchikov \cite{CZ}. The same kind of argument was used in the study of the
Friedel oscillations in \cite{LeclairLS}. There, it was shown that 
that
\begin{equation}
\label{Factorisation}
 g_T=\langle \exp\left[\pm{i\over
2}(\phi_{L}+\phi_{R})(x,y)\right]\rangle=
\langle\sigma(x)\rangle_{h}\langle\sigma(x)\rangle_{\infty}
\end{equation}
where $\langle\sigma\rangle_{h}$ denotes the one point function of the spin
operator in the Ising model with boundary field $h\propto v$ 
(the right hand side does not depend on $y$ when the two
vertex operators are inserted at the same (imaginary) time), and $x$ is the
distance from the boundary. This is similar but different from the usual result in the bulk 
\cite{ItzZub,Schroer}
\begin{eqnarray}
\left<\sigma(z,\bar{z})\sigma(0,0)\right>^{2}&=&\left<
0|\sin(\Phi(z,\bar{z})/2)\sin(\Phi(0)/2))|0\right>\nonumber\\
\left<\mu(z,\bar{z})\mu(0,0)\right>^{2}&=&\left<
0|\cos(\Phi(z,\bar{z})/2)\cos(\Phi(0)/2))|0\right> .
\end{eqnarray}

This result immediately extends to the case where
the right and the left field are inserted at different points, since 
the  correlator depends on $x_{2}-x_{1}+i(y_{2}-y_{1})$ only: 
\begin{eqnarray}
 g_T&=&\langle\exp(-i\phi_{L}(-x_2,y_2)/2)\exp(-i\phi_R(x_1,y_1)/2)\rangle 
\\ \nonumber &=&\langle
\sigma((x_{2}-x_{1}+i(y_{2}-y_{1}))/2\rangle_{h}
\langle\sigma((x_{2}-x_{1}+i(y_{2}-y_{1}))/2)\rangle_{\infty}
\end{eqnarray}
Using  form factors, one can expand the one point function of
the Ising spin operator as 
 \begin{equation}
\langle\sigma(x)\rangle_{h}=\sum_{n=0}^{\infty}{1\over n!}
\int_{-\infty}^{\infty}\prod_{i=1}^{n}\left\{{d\beta_{i}\over
2\pi} \tanh {\beta_{B}-\beta
_i\over
2}e^{-2xe^{\beta_{i}}}\right\}
\prod_{i<j}\left(\tanh{\beta_{i}-\beta_{j}\over 2}\right)^{2}
\end{equation}
and one can also check the factorization (\ref{Factorisation})
of $g$ directly using the form-factors expansion as described in
\cite{LeclairLS}.The case of infinite magnetic field corresponds to $h\rightarrow \infty$
and since $p_B=e^{\beta_B}=4\pi h^2$ it also corresponds to $p_B\rightarrow \infty$
in this formula. In that case the result is known to be $1/x^{1/8}$ from boundary conformal field 
theory.
Recall
on the other hand the result that
\begin{equation}
    \langle\sigma(x)\rangle_{h}\propto~ x^{3/8} \sqrt{\frac{p_B}{4\pi}} ~e^{p_B x}~K_{0}( p_B
    x)\label{nicezam}
    \end{equation}
one thus find that 
\begin{equation}
G_{T}(1,2)\propto~ \sqrt{\frac{p_B}{4\pi}}~e^{p_B x/2}~K_{0}(p_B
x/2),~~~x\equiv x_{2}-x_{1}+i(y_{2}-y_{1})
\end{equation}
where the factor $1/2$ came up in the argument of the
Bessel function since in the initial Ising case, our variable $x$
would be twice the value in  (\ref{nicezam}). We can now use the integral representation
$$
K_{0}(z)={e^{-z}\over\sqrt{2}}\int_0^{\infty}
{e^{-zt}\over \sqrt{t}}\left(1+{t\over 2}\right)^{-1/2}dt
$$
to write 
\begin{equation}
G_{T}(1,2)\propto~ \int_{0}^{\infty} e^{-xp}p^{-1/2}\left(1+{p\over
  p_{B}}\right)^{-1/2}~dp
\end{equation}
giving rise to the stunningly simple momentum dependent transmission amplitude
\begin{equation}
G_{T}(p)\propto~p^{-1/2}~\left(1+{p\over
 p_{B}}\right)^{-1/2}\label{maini} .
\end{equation}
Finally, we perform a shift (in the rapidity variable) of the integration contour to get back to
the theory in real time, quantized with $x$ as the space axis, and the interaction at
the origin, which amounts to $p\rightarrow ip$: 
\begin{equation}
G_{T}(p)\propto~p^{-1/2}~\left(1+{ip\over
 p_{B}}\right)^{-1/2} .
 \label{mainii}
\end{equation}
Note that this result could also be obtained  by
quantizing the theory in the crossed channel. In that case, the
boundary interaction appears because asymptotic states are of the
form $|\beta\rangle_{R}+R(\beta)|\beta\rangle_{L}$, and leads to
expressions with the shifted contour, and $K$  replaced by $R$. This 
correspondence is
discussed in details in \cite{LSS}.

At low energy, $G_{T}\propto p^{-1/2}(1+\hbox{cst}\frac{p}{p_{B}})$ and at
high energies $G_{T}\propto 
p^{-1/2}\sqrt{\frac{p_{B}}{p}}$: this is the expected behaviour. Indeed,
in the
 limit of small $p_{B}$ the perturbative expansion should be linear in
 $v\propto \sqrt{p_{B}}$, and one thus expects the
 expansion to go as $\sqrt{p_{B}/p}$. On the other hand, at large $p_{B}$, one
 approaches the IR  fixed point (total transmission), along the
 operators
 $\cos \tilde{\Phi}$  of dimension $d=2$, 
 and $\left(\partial\tilde{\Phi}\right)^2$ \cite{LesageSaleur}(here, 
 $\tilde{\Phi}=\phi_{R}-\phi_{L}$ is the dual of  the
 field $\phi$, and the theory is defined on the half line). The
 coupling of these operators must
 therefore have dimension $-1$, and thus be proportional to
 $v^{-2}$ since $v$ has dimension $1/2$. 
 The perturbative series starts linearly
 in the coupling, that is linear in $p/p_{B}$, as we just found.
    
 It is interesting to recall that the Bessel function $K_{0}$
 only admits an {\sl asymptotic expansion}  in powers of the argument
 $$
 e^{z}K_{0}(z)\approx\sqrt{\pi\over 2z}\left[
 \sum_{k=0}^\infty {1\over (2z)^{k}}{\Gamma(k+1/2)\over
 k!\Gamma(-k+1/2)}\right] .
 $$
 In our problem, this
 corresponds to the expansion near the IR fixed point (low energy of 
 incident particles,
 large $p_{B}$), which is thus only asymptotic for the correlator (in
 fact, the expansion of $K_{0}$ is good only up to exponential terms 
 $e^{-z}$. Here this corresponds to terms $e^{-x/(1/v^{2})}$,
 which are non perturbative in term of the IR coupling constant
 $1/v^{2}$). All these difficulties disappear however
in  the momentum dependent amplitude, whose expansion 
 is now convergent, thanks to 
  the Laplace
 transform.  
 
 Let us get back finally  to the expression for $
 \langle\sigma(x)\rangle_{h}$.
 The leading order in $h$ is proportional to $hx^{3/8}\ln (xh^{2})$. 
 In perturbation theory meanwhile, the leading order involves an
 integral over the boundary (in the Euclidian formalism) $\int dy {1\over
 \left(x^{2}+y^{2}\right)^{1/2}}$ which must be regularized by the
 introduction of an IR cut-off. The situation is similar to the massive bulk Ising model or the bulk sine-Gordon model at $\beta^2=4\pi$ (see \cite{Konik}
and referecnes therein). An additional renormalization is required, leading to the replacement 
of the cut-off by the inverse of the mass - here, by the inverse coupling constant
 $1/p_{B}$. 
Alternatively, one could instead consider the 
correlator at finite temperature. $1/T$ then acts
 as an IR cut-off, and the scaling limit of $\langle\sigma\rangle$ is entirely
 well defined. Taking the limit $T\rightarrow 0$ then gives the
 result mentioned above \cite{LeclairLS}.

\subsection{The reflection amplitude}

The  reflection amplitude is an intriguing quantity. Let us think of 
it 
 first by  considering  the vicinity of the IR fixed
 point, which is  is approached along the operator
 $\cos\tilde{\Phi}$. In an IR perturbation
 theory, the average of the quantity
 $\exp\left[i(\phi_{R}-\phi_{L})/2\right]$ will vanish to every
 order, because a charge neutral exponential can never be formed.
 Therefore, if the reflection amplitude happpened to be non zero, if
 would be entirely due to non perturbative effects! In fact, we think
 this amplitude is exactly zero.
 This can be argued more solidly by considering  the UV point of view.
 The form of the IR perturbation shows that, right at the  UV fixed
 point, the field at the boundary
 obeys $\phi_{R}-\phi_{L}(0)=(2n+1)\pi$, $n$ an
 integer. It follows
 that the quantity $\exp\left[i(\phi_{R}-\phi_{L})/2\right]$ changes 
 sign when one goes from one well of the boundary potential 
to the next. Since a proper
 conformal invariant boundary condition involves a sum over all
 wells, the average
 $\langle\exp\left[i(\phi_{R}-\phi_{L})/2\right]\rangle$
 has to vanish right at the fixed point. 
 This observation can be made more formal by appealing to the
  formalism of conformal boundary conditions (notations follow those
  of \cite{houches}; $\alpha_n$ are the usual bosonic modes; 
$|w,k>$ the zero modes). The Dirichlet state is
  given by consideration of the IR fixed point (radius
  $r={1\over\sqrt{\pi}}$)
  \begeq
  \left|B_D(\Phi_0)\right>= {1\over\sqrt{2r\sqrt{\pi}}} \sum_{k=-\infty}^\infty e^{-ik\Phi_0/r} 
  \exp\left[-\sum_{n=1}^\infty {\alpha_{-n}\bar{\alpha}_{-n}\over n}\right]\left|0,k\right>
  \endeq
 The Neumann state compatible with this is 
 \begeq
 \left|B_N(\tilde{\Phi}_0)\right>={1\over 2}\sqrt{2r\sqrt{\pi}}
 \sum_{w=-\infty}^\infty
 e^{-2i\pi rw\tilde{\Phi}_0}\exp\left[\sum_{n=1}^\infty {\alpha_{-n}
 \bar{\alpha}_{-n}\over n}\right]
 \left|w,0\right>
 \endeq
 While the operator $\exp\left[i(\phi_{L}-\phi_{R})\right]$
 corresponds to $w=1$, and hence can have a non trivial one point function
 with $N$ boundary conditions, the operator 
 $\exp\left[i(\phi_{L}-\phi_{R})/2\right]$ would correspond formally
 to $w=1/2$. The corresponding state is not in the decomposition of
 the $N$ boundary state, and therefore the one point function has to vanish. 

In a UV perturbation theory,
the same sum over wells (which is compatible with the perturbation) 
will give rise to the same result.
(Notice how the situation for the quantity 
$\exp\left[i(\phi_{R}+\phi_{L})/2\right]$ is entirely different. In 
the UV, because the perturbation is
$\cos\Phi/2=\cos(\phi_{R}+\phi_{L})/2$, neutral combinations can be
formed, and one finds non trivial perturbative contributions. In the
IR, the field obeys now $(\phi_{R}+\phi_{L})/2=(2n+1)\pi$, and thus 
the exponential of interest is invariant when one goes from one well
to the next.)

Finally, the vanishing  can also be understood in terms of
 form-factors. In the massive
theory, the operator $\exp\left[i(\phi_{L}-\phi_{R})/2\right]$ is not
neutral, and increases the $L-R$ charge by one unit. The  non zero 
form-factors therefore will have to involve an {\sl odd} number of
particles. Since the boundary state is a superposition of an even
number of particles only, the one point function of this operator
must vanish identically. A close correspondence with the disorder
operator in the Ising model is discussed in the appendix. 

Of course, right at the UV fixed point, one can ``by hand'' restrict to one well
of the potential, and find a non zero amplitude, of the expected form
$G_{R}(p)\propto p^{-1/2}$. It is tempting to force this amplitude
to be non zero away from the fixed point as well.
%
 %

To do so, one can again use  the form-factors
approach. As explained in the appendix, the non zero value right at
the UV fixed point can be obtained by adding to the boundary state a 
particle of vanishing energy (and momentum). It is reasonable to
try to define a non zero amplitude  near  the 
UV fixed point  by generalizing 
the addition of this particle at finite value of the coupling. This
leads to the result 
\begin{equation}
  g_{R}=\langle\exp(i\phi_{L}(-x_2,y_2)/2)\exp(-i\phi_R(x_1,y_1)/2)\rangle=
{1\over (x_{2}-x_{1}+i(y_{2}-y_{1}))^{1/8}}
\langle\sigma[(x_{2}-x_{1}+i(y_{2}-y_{1})/2]\rangle_{\tilde{h}}
    \end{equation}
where the second quantity has the form factor expansion
 \begin{equation}
     \langle\sigma(x)\rangle_{\tilde{h}}\equiv\sum_{n=0}^{\infty}{(-1)^{n}\over n!}
     \int_{-\infty}^{\infty}\prod_{i=1}^{n}\left\{{d\beta_{i}\over
     2\pi} \tanh {\beta_{B}-\beta\over
     2}e^{-2xe^{\beta_{i}}}\right\}
     \prod_{i<j}\left(\tanh{\beta_{i}-\beta_{j}\over
     2}\right)^{2}\label{howpuzzling}
     \end{equation}
and the minus signs have been generated by the introduction of the
 particle at zero energy. 
 
 Of course, this expansion is once again divergent. Moreover, in
 contrast with the case of the transmission amplitude, we do not even
 know the value in the IR, so we cannot divide by it to form a finite
 quantity. 
 We can however observe  that the logarithmic derivative of the
 correlator is regularizable by the same procedure previously used
 for $\langle\sigma\rangle_{h}$. The first two terms read for instance
 \begin{eqnarray}
     {\partial\over\partial\beta_{B}}
     \ln\langle\sigma(x,p_{B})\rangle_{\tilde{h}}=&-&\int_{0}^{\infty} {du\over\pi}
     {p_{B}\over (p_{B}+u)^{2}}e^{-2xu}\nonumber\\
     &-&\int_0^\infty {dudv\over \pi^{2}} (u+v)p_{B} {p_{B}^{2}-uv\over
     (p_{B}+u)^{2}(p_{B}+v)^{2}}e^{-2x(u+v)}+\ldots
     \end{eqnarray}
 (where recall $p_{B}=e^{\beta_{B}}$). The integrals converge now at the origin, including when $p_{B}=0$. 
 Call the series in the right hand side $\Sigma(xp_{B})$. We can thus 
 try to write 
 $$
 \langle\sigma(x,p_{B})\rangle_{\tilde{h}}={1\over x^{1/8}}
     \exp\left[\int_{0}^{p_{B}}{dp'_{B}\over p'_{B}}~
     \Sigma(xp'_{B})\right]
    $$
Although right at $p_{B}=0$ this expression reproduces the right
result, there is a problem as soon as $p_{B}\neq 0$. This is because, 
at every order we have explored, $\Sigma(z)$ goes to a finite
quantity as $z\rightarrow 0$, and thus the integral diverges at the
origin! This can be studied in more details 
at the first order already, where the exponential reads, for small
$p_{B}$, 
$$
 \exp\left[-\int_{0}^{p_{B}} {dp_{B}'\over p'_{B}}{1\over\pi}\right]
$$
To make the integral finite we have to introduce a new (IR) cut-off
$\epsilon$, with 
dimension of temperature $=\hbox{length}^{-1}$, so we get 
$$
\exp\left(-{1\over\pi}\ln(p_{B}/\epsilon)\right)=(\epsilon/p_{B})^{1/\pi}
$$
If we now let the cut-off go to zero, we see that the amplitude again
vanishes. A similar result was noticed independently by N. Sandler (private communication). 

This result is confirmed by higher order calculations. We think it is
a consequence of the ill defined nature of the reflection amplitude
within field theory, and can be interpreted by saying that this
reflection amplitude is identically zero for any $p_{B}\neq 0$, in
the scaling limit. The only way to define this amplitude is to keep an
additional (IR) cut-off, and then, to first order in the form-factors
expansion, one finds
\begin{equation}
    \langle\sigma(x)\rangle_{\tilde{h}}={1\over x^{1/8}} \left({\epsilon\over
    p_{B}}\right)^{1/\pi}\exp\left[{1\over\pi}
    \int_{0}^\infty {dz\over z(1+z)}
    \left(e^{-2zxp_{B}}-1\right)\right]
    \end{equation}
 It is in fact possible to obtain the
 exact value of the one point function by using more results about
 the Ising model, as detailed in the appendix:
 \begin{equation}
     \langle\sigma(x)\rangle_{\tilde{h}}={1\over x^{1/8}} \left(\epsilon x
    \right)^{1/2}\sqrt{2p_{B}x}\left(2\Psi'-\Psi\right)\label{brave}
    \end{equation}
 where again $\Psi=\Psi(1/2,1;z)={e^{z/2}\over\sqrt{\pi}}K_0(z/2),~z=2p_{B}x$. From this the
 result follows that $\langle\sigma\rangle_{\tilde{h}}\propto
 \epsilon^{1/2}x^{3/8}$ at large values of $p_{B}x$. 
 
 To each order we have explored, the value of the exponent at the origin 
 was
 ${1\over \pi}$, that is, every contribution but the first one seemed 
 to vanish exactly. Meanwhile, the exact value expected from the
 formula (\ref{brave}) is $1/2$. Most likely what happens is that the
 form-factors expansion does not converge uniformly at the origin -
 an identical feature takes place in the case of the ordinary one
 point function $\langle\sigma\rangle_{h}$.

Going back now to the physical amplitude we find 
\begin{equation}
G_{R}\propto{1\over x^{1/2}}\left(\epsilon x
\right)^{1/2}\sqrt{2p_{B}x}\left(2\Psi'-\Psi\right)\label{compconj}
\end{equation}
In particular it follows that $G_{R}(p)\propto
p^{-1/2}\left(\epsilon/p\right)^{1/2}$ in the IR 
limit.

\subsection{Other amplitudes}

What one may call an ``Andreev process'' meanwhile would
correspond to a Laughlin quasiparticle incident on the point
contact giving rise to an electron
 in the same chiral
Luttinger liquid, and  a Laughlin quasihole
in the other chiral Luttinger
liquid. The amplitude for this process is 
\begin{equation}
    G_{A}=\left<e^{-i\phi_{2}(x_{3}>0,t_{3})/\sqrt{2}}
    e^{i\sqrt{2}\phi_{1}(x_{2}>0,t_{2})}
    e^{-i\phi_{1}(x_{1}<0,t_{1})
    /\sqrt{2}}
    \right>
   \end{equation}
  The interacting part of this correlator, after using even odd
  combinations and folding, becomes 
  \begin{equation}
      g_{A}=\left<e^{i\phi_{L}/2}e^{-i\phi_{R}/2}
      e^{i\phi_{L}}\right>
      \end{equation}
  and the argument has the form ${\phi_{L}+\phi_{R}\over
  2}+\phi_{L}-\phi_{R}$. This time, the exponential is invariant when
  going from one well to the next, so the average will be non trivial.
  The amplitude vanishes both  in the UV and IR limits, but has non
  vanishing perturbative contributions in the vicinity of either
  fixed point. All indications are that this amplitude is well
  defined in the scaling limit.
   
In fact, the momentum dependent amplitude $G_{A}(p,q)$ (with say $q$ 
the amplitude of the outgoing electron)  has the same dimension as
the would be amplitude for the reflection process (\ref{compconj}). It is likely  that
the regularized reflection amplitude we have defined previously (\ref{compconj})
coincides with the limit of $G_{A}(p,q)$ as $q$ goes to zero (and
then $q$ has to be identified with $\epsilon$). 
       
\begin{figure}
 \vspace{0.3cm}
 \begin{center}
 \epsfig{file=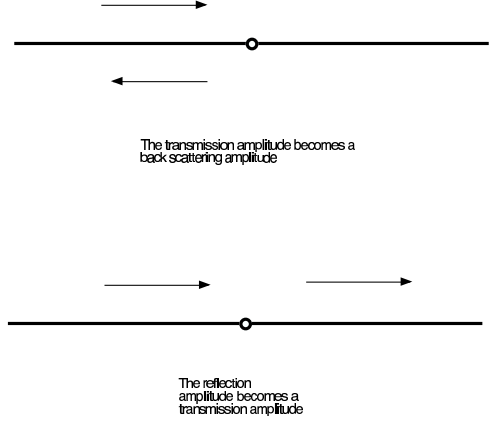,scale=0.4}  
 \caption{\label{fig2} Tunneling between two identical chiral  Luttinger
 liquids mapped onto the problem  of a single impurity within a non
chiral liquid.}
 \end{center}
 \end{figure}

Note finally that by folding one can  map the geometry of two  R moving Luttinger liquids
to  the geometry of a single non chiral Luttinger liquid. This can actually be done in several ways: the most natural  is illustrated 
on the figure \ref{fig2}. In
this case, what we called before transmission amplitude becomes a 
backscattering amplitude (small at small $v$) and what we called
reflection amplitude becomes transmission amplitude (small at
large $v$).

\section{Point contact between a Fermi and a Luttinger liquid}

\subsection{Set-up}

\begin{figure}
\vspace{0.3cm}
\begin{center}
\epsfig{file=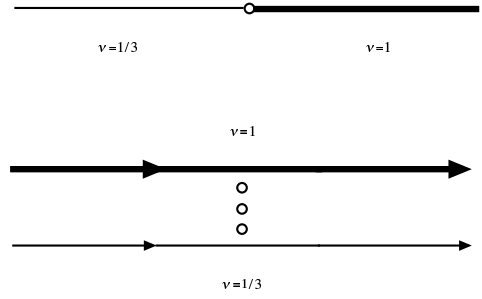,scale=0.4}  
\caption{\label{fig3} Tunneling between two different chiral  Luttinger
liquids mapped onto the problem of tunneling between two
different non chiral Luttinger liquids.}
\end{center}
\end{figure}

A fundamental component in the experimental study of Luttinger
liquids is the presence of a point contact with a three dimensional reservoir
(in fact, a semi infinite three dimensional reservoir, whose boundary
is a plane \cite{CF}).
The analysis of the situation is 
somewhat similar to the Kondo problem. Electrons in the  3d reservoir
can be organized not into plane waves but into spherical waves, and
one finds that only the $s=0$ wave interacts with the Luttinger
liquid (the Kondo impurity in the other case). Because there is a
single mode, it can be considered as well as arising from a one
dimensional conductor, and one can reduce the problem to the one of a
point contact between a 1D Fermi liquid and a 1D Luttinger liquid. The situation then looks as
in the figure  \ref{fig3}. 

By unfolding, one can map this problem onto the one of two chiral (say
right moving)
liquids with an interaction localized at the origin, corresponding to
the bottom part of \ref{fig3}. 
The translation between the two is obvious (and different from the folding in figure \ref{fig2}): for instance, a process
where an electron comes in and bounces within the Fermi liquid
corresponds in the unfolded picture to an electron going through on
the top line.

In a particular experiment,
different tunneling terms will be induced by the realization of the 
point contact. At low energy - or low temperature -
the decoupled fixed point, where the two electronic liquids are 
decoupled, is stable. 
To leading order, the tunneling that kicks in as energy is increased  
is a process where 
one electron from the Fermi liquid is exchanged for an ``edge'' 
electron in the Luttinger liquid.  The amplitude for this term
is traditionnally called $\Gamma$.  
In the limit of small $\Gamma$, the Lagrangian reads
\begin{equation}
L={1\over 4\pi} \partial_x\phi_a(\partial_t-\partial_x)\phi_a+{1\over 
4\pi} \partial_x\phi_b(\partial_t-\partial_x)\phi_b
+\Gamma\delta(x)\cos[{1\over\sqrt{\nu}}\phi_a-\phi_b],
\end{equation}
where the field $\phi_a$ is associated with the Luttinger liquid, and 
$\phi_b$ with the Fermi liquid. This Lagrangian can be brought in a 
simpler form by the definition of new rotated fields $\varphi_{a,b}$:
\begin{equation}
L={1\over 4\pi} 
\partial_x\varphi_a(\partial_t-\partial_x)\varphi_a+{1\over 4\pi} 
\partial_x\varphi_b(\partial_t-\partial_x)\varphi_b
+\Gamma\delta(x)\cos 
\left[{1\over\sqrt{g'}}(\varphi_a-\varphi_b)\right]
\end{equation}
where 
$$
{1\over g'}={\nu+1\over 2\nu}
$$
We will restrict here to  $\nu={1\over 3}$,  which gives rise to 
$g'={1\over 2}$. The fields can now be gathered into an even 
combination
which is free, and an odd combination that sees a boundary 
sine-Gordon type interaction.

The perturbation induced by the tunneling term  has dimension $2$, 
and is {\sl irrelevant}. This means that at low energies, there is
essentially no tunneling between the initial Fermi and Luttinger
liquids: one is at the fixed point where transmission is zero. 
This  crucial feature of  the problem, 
induces considerable complications. Indeed, suppose  one were to 
consider an experimental realization
starting from the trivial situation of decoupled wires at low energy, 
and increasing the energy to explore the non trivial scattering 
processed. In term of renormalization group, this would translate 
into going 
{\sl against the RG flow}, since the low energy fixed point is 
stable, and the coupling between the wires is an irrelevant 
perturbation.  
As is well known, such a situation does not have a 
well defined universal limit, as  the RG  trajectory is influenced by 
the many other irrelevant operators 
present in the problem. In our case, this means for instance that one 
also has to take into account the density density coupling, which 
also has dimension $2$, as well as other fields maybe, and the 
results will be affected by the balance of these different terms. A
priori, only extreme fine tuning would allow one to reach the perfect
transmitting fixed point at high energy.

Fortunately, there seems to be a way to  realize the high energy fixed 
point $\Gamma=\infty$.  Indeed, 
in a very interesting paper,   C. Kane \cite{Kane} pointed out that if tunneling
between the two liquids takes place through an impurity, it is
possible, by tuning two parameters only, to achieve perfect resonance that is, to have
perfect tunneling between the Fermi and
Luttinger liquids. 
Away from the resonance, deviations are controlled by the single
relevant operator at the resonant fixed point, and the problem
becomes formally identical  to the by now well known problem of edge
states tunneling at $\nu=g'$. 

\begin{figure}
\vspace{0.3cm}
\begin{center}
\epsfig{file=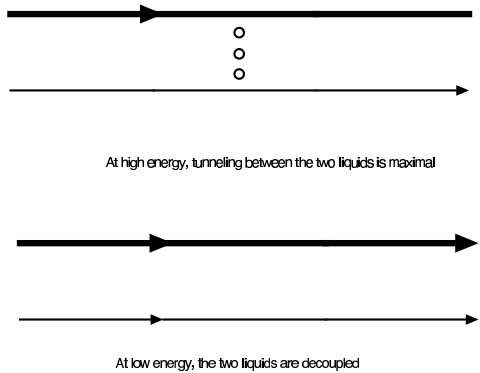,scale=0.4}  
\caption{\label{fig4} Tunneling between the Fermi liquid and the
$\nu=1/3$ 
Luttinger liquid as a function of the energy}
\end{center}
\end{figure}

In this new situation, 
at high energy, there is maximum tunneling between the 
Fermi and the Luttinger liquid which are strongly coupled, while at
low energy, there is no remaining tunneling, the two are decoupled, 
and the scattering is entirely diagonal. This is illustrated on the
figure \ref{fig4}.


The problem in the vicinity of the perfectly transmitting fixed point 
can be formulated in terms of dual fields, with Lagrangian
\begin{equation}
L={1\over 4\pi} 
\partial_x\tilde{\varphi}_+(\partial_t-\partial_x)\tilde{\varphi}_++{1\over 
4\pi} \partial_x\tilde{\varphi}_-(\partial_t-\partial_x)
\tilde{\varphi}_-
+\tilde{\Gamma}\delta(x)\cos \sqrt{2g'}\tilde{\varphi}_-
\end{equation}
and $\tilde{\Gamma}\propto \Gamma^{-1/2}$. In this case, the perturbation is relevant, with dimension $1/2$. 
As energy is lowered, physical processes where electrons and Laughlin 
quasiparticles are incident on the point contact and scatter into
various combinations of outgoing electrons and Laughlin 
quasiparticles will have energy dependent amplitudes. 

In the following we denote the field $\tilde{\varphi}_-$ simply by 
$\phi$. We will also denote $\tilde{\varphi}_+$ by $\phi'$.
We fold the model so that instead of having right movers
on the full line, we have left and right movers on the half line only,
with the Lagrangian
\begin{equation}
L={1\over 8\pi}\int_{-\infty}^0 dx 
\left[(\partial_x\Phi)^2-(\partial_t\Phi)^2\right]+\tilde{\Gamma}\cos{1\over 
2}\Phi(0)
\end{equation}
In states are now right moving, and out states are left moving. As 
before, the perturbation has dimension $1/2$, and is relevant. 

\begin{figure}
\vspace{0.3cm}
\begin{center}
\epsfig{file=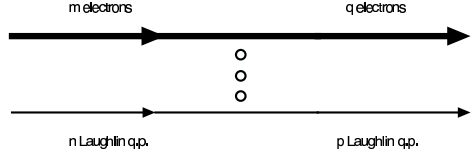,scale=0.4}  
\caption{\label{fig5} A process where $m$ electrons and $n$ Laughlin 
quasiparticles scatter into $q$ electrons and $p$ Laughlin
quasiparticles.}
\end{center}
\end{figure}

The most general quantities  we are interested in read, in the
original $\phi_{a,b}$ variables, 
\begin{equation}
    G(4,3,2,1)=\langle T~ e^{i\sqrt{\nu}p\phi_{a}^{out}(4)}
    e^{iq\phi_{b}^{out}(3)}
    e^{-i\sqrt{\nu}n\phi_{a}^{in}(2)}
    e^{-im\phi_{b}^{in}(1)}\rangle
\end{equation}
and are related (a precise definition would require, in particular, a choice of massless basis) to the amplitude for $m$ electrons and $n$ Laughlin 
quasiparticles to scatter into $q$ electrons and $p$ Laughlin 
quasiparticles, as represented on figure \ref{fig5}.

Following through the transformations detailed previously,
this translates into
\begin{eqnarray}
G(4,3,2,1)=\langle T~
\exp i\left[-{p\over 2}\phi_L(4)+{q\over 
2}\phi_L(3)-{n\over 2}\phi_R(2)+{m\over 2}
\phi_R(1)\right]\times\nonumber\\
\exp i\left[{p\over 2\sqrt{3}}\phi'_L(4)+{q\sqrt{3}\over 
2}\phi'_L(3)-{n\over 2\sqrt{3}}\phi'_R(2)-m{\sqrt{3}\over 
2}\phi'_R(1)\right]\rangle
\end{eqnarray}

The fact that the $\phi'$ field is non interacting and sees Neumann 
boundary conditions imposes the (charge neutrality) constraint
\begin{equation}
{p\over 3}+q={n\over 3}+m
\end{equation}
The full correlator will vanish in the limit of decoupled wires 
$\tilde{\Gamma}=\infty$ ($\Gamma=0$) unlesss $n-m=p-q$ 
and thus, $q=m,p=n$. This means that the scattering is 
diagonal,  and that electrons and Laughlin quasiparticles go through 
without scattering, the well understood low energy fixed point. The 
correlator  will vanish in 
the limit of totally coupled wires (perfect tunneling) $\tilde{\Gamma}=0$ 
($\Gamma=\infty$) unless $n-m=-(p-q)$.  
In this limit  we have then, combining the two conservation equations,
\begin{eqnarray}
q={1\over 2} m+{1\over 2}n\nonumber\\
p={3\over 2} m-{1\over 2} n
\end{eqnarray}
Note that at the high energy fixed point, 
strange things occur, as discussed in \cite{SCW}. For instance,
all the processes
where a single electron comes in $m=1,n=0$, have vanishing amplitude!
Still, probability must be conserved - so what may happen is 
that an infinity of processes have non vanishing amplitudes at non
zero $\tilde{\Gamma}$, with the sum of the associated probabilities
equal to unity, and that the corresponding series is not uniformly
convergent: as $\tilde{\Gamma}\rightarrow0$, each individual term
vanishes, but the sum remains equal to 1.  We do not know how to
explore this situation in more detail. 

For simplicity, we will mostly restrict to  correlators of the form 
\begin{eqnarray}
G(3,1)&=&g(3,1)~\langle
T~\exp(i\sqrt{3}\alpha\phi'_L(-x_3,t_3))\exp(-i\sqrt{3}\alpha\phi'_R(x_1,t_1))\rangle\nonumber\\
&=&g(3,1) ~{1\over (x_3-x_{1}-t_{3}+t_1+i\epsilon~sign(t_{3}-t_{1}))^{3\alpha^2}},
\end{eqnarray}
where $x_{3}>0,x_{1}<0$, and 
\begin{equation}
g(3,1)=\langle
T~\exp(i\alpha\phi_L(-x_3,t_3))\exp(i\alpha\phi_R(x_1,t_1))\rangle.
\end{equation}
These correlators correspond either to 
$p=n=0,q=m$ or $q=m=0,p=n$ and describe the scattering of $q$
electrons into $q$ electrons, or $p$ Laughlin quasiparticles into $p$
Laughlin quasiparticles.

In the limit $\tilde{\Gamma}=0$ (high energy fixed point), the field $\phi$ sees Neumann 
boundary conditions, and $g$ vanishes by charge neutrality. This 
corresponds to 
the Fermi liquid and 
the Luttinger liquid being strongly coupled, so that all processes of 
finite amplitude involve non diagonal scattering, like 
electrons scattering into Laughlin 
quasiparticles. 
In the limit $\tilde{\Gamma}\rightarrow\infty$, on the contrary, the 
field sees Dirichlet boundary conditions, the scattering of electrons 
or Laughlin quasiparticles 
is diagonal,
and the correlator takes a finite value, 
\begin{equation}
g(3,1)\rightarrow {1\over 
(x_3-x_1-t_3+t_1+i\epsilon~sign(t_{3}-t_{1}))^{\alpha^2}},~~~\tilde{\Gamma}\rightarrow\infty
\end{equation}
In general, this correlator will exhibit a cross-over behaviour, 
which we will 
determine using the form-factors technique. Knowledge of the 
correlators will then allow us to determine the amplitude of 
interest.

\subsection{The case $2e\rightarrow 2e$}

The case $|\alpha=1|$ is  special, and much simpler. Indeed, as
discussed in the first part, the 
operator $V_\pm$ behaves like a  fermion field,
$V_{\pm~1}\propto \psi^{(\dagger)}$ in a free fermion theory, the $-$ 
soliton like a fundamental fermion,  the $+$ soliton like a hole, 
\begin{eqnarray}
<0|\psi_L|\beta>_{L-}=\mu  e^{\beta\over 2} e^{-ie^\beta 
(t+x)};~~~_{L+}<\beta|\psi_L|0>=  \mu e^{\beta\over 2} e^{i e^\beta 
(t+x)} \nonumber\\
<0|\psi_R|\beta>_{R-}=\mu  e^{\beta\over 2} e^{-ie^\beta 
(t-x)};~~~_{R+}<\beta|\psi_R|0>= \mu  e^{\beta\over 2} e^{i e^\beta 
(t-x)} \nonumber\\
\end{eqnarray}
where $\mu$ is a normalization factor ($\mu=\sqrt{2\pi}$ if one wants
the fermions two point function to be normalized to unity). Similarly, for the dagger 
operators,
\begin{eqnarray}
<0|\psi^\dagger_L|\beta>_{L+}=\mu  e^{\beta\over 2} e^{-ie^\beta 
(t+x)};~~~_{L-}<\beta|\psi_L^\dagger|0>=  
\mu
e^{\beta\over 2} e^{i e^\beta (t+x)} \nonumber\\
<0|\psi_R^\dagger|\beta>_{R+}=\mu  e^{\beta\over 2} e^{-ie^\beta 
(t-x)};~~~_{R-}<\beta|\psi_R^\dagger|0>= 
\mu  e^{\beta\over 2} e^{i e^\beta (t-x)} \nonumber\\
\end{eqnarray}

It follows immediately that, for $\alpha=1$, using the same
technique as in the first part, 
\begin{equation}
g(3,1)=<0|\psi_L(3)\psi^{\dagger}_R(1)|B>\equiv
<\psi_L(3)\psi^\dagger_R(1)>\propto \int_{-\infty}^\infty 
{e^\beta\over 1+e^{\beta-\beta_B}}{d\beta\over 2\pi} 
e^{e^\beta(x_1-x_3+i(y_{1}-y_{3}))}
\end{equation}
Writing now the integral representation
\begin{equation}
G(3,1)=\int_0^\infty dp~~ G(p,p_B) e^{p(x_1-x_3+i(y_{1}-y_{3}))}
\end{equation}
we arrive, after convolution,  at the following formula
\begin{equation}
G(p,p_B)\propto  -{ p_B\over 2}
\left[(p+p_B)^2\ln(1+p/p_B)-{3p^2+2pp_B\over 2}\right]
\end{equation}
such that 
\begin{eqnarray}
G(p,p_B)&\approx& - {p^3\over 6},~~~{p\over 
p_B}\rightarrow 0\nonumber\\
G(p,p_B)&\approx& -{1\over 2} p^2 p_B\ln p_B,~~~{p\over 
p_B}\rightarrow\infty
\end{eqnarray}
where we recall that the high energy limit is $p_B\rightarrow 0, 
\tilde{\Gamma}\rightarrow 0$.

Physically, the case $\alpha=1$ corresponds to $q=m=2$, ie a process 
where a wave packet of charge $2e$ comes from the left in the 
Fermi liquid, and goes accross to a charge $2e$ wave packet still in 
the Fermi liquid. In the limit $\tilde{\Gamma}=\infty$, this process
occurs with probability $1$, and it occurs with probability $0$ in 
the limit $\tilde{\Gamma}=0$.

\begin{figure}
\vspace{0.3cm}
\begin{center}
\epsfig{file=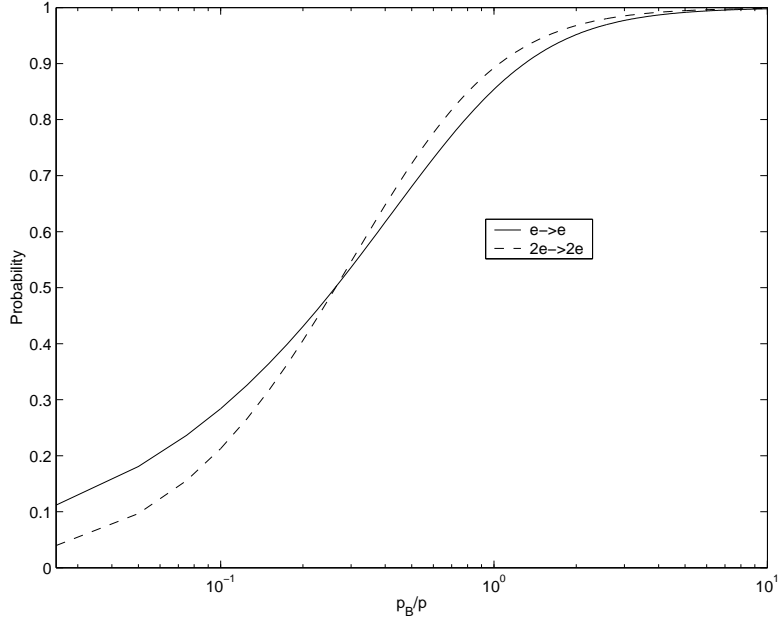,scale=0.6}  
\caption{\label{figProb} Probabilities for the processes $2e\rightarrow 2e$ and
$e\rightarrow e$.}
\end{center}
\end{figure}

The meaning of $G$ as a probability in between these two limits 
requires some more discussion. The problem boils down to what one
wishes to call a wave packet of charge $2e$. In the bosonized theory,
it is convenient to think of this as the object created by
$e^{i\phi_{L,R}}$, which in the original Fermi liquid corresponds to 
the quantity $\chi\partial\chi$ (the fermion $\chi$ is the fermion
field in the Fermi liquid, and has no direct relation with the
fermionic field $\psi$ which appears in the reformulation of the
boundary sine-Gordon theory at the value $\beta^{2}=4\pi$).  When
written in terms of modes, $\chi\partial\chi$ reads as an
superposition of states with two electrons at different momenta,
weighed by some energy dependent coefficient. The corresponding
creation operators have commutation relations of the form
$[a^{\dagger}(k),a(k')]\propto k^{3}\delta(k-k')$, and it is
convenient to use these states as normalized states for charge $2e$
wave packets. With this convention,
the probability that such a wave packet goes through unaltered  is
\begin{equation}
P_{2\rightarrow 2}=\left|{G(ip,p_B)\over 
G(ip,\infty)}\right|^2=9\left|ix(1+ix)^2\ln\left(1+{1\over 
ix}\right)-{3ix\over 2}-(ix)^2\right|^2
\end{equation}
with $x={p_B\over p}$. This probability is represented in figure \ref{figProb}.

\subsection{The case $e\rightarrow e$}

The case $\alpha={1\over 2}$, which corresponds to a process 
where an electron comes out as an electron, is of course much more complicated to
study, and bears close resemblance to the transmission amplitude of
the first part. 
A closed form expression is again possible, and one finds that, in
imaginary time,
\begin{equation}
G(3,1)\propto x^{-1/2}\sqrt{\frac{p_B}{4\pi}}e^{p_B x/2}K_{0}(p_B
x/2),~~~x=x_{3}-x_{1}+i(y_{3}-y_{1})
\end{equation}
This provides a closed form expression for the Laplace transform
\begin{equation}
    G(p/p_{B})={1\over\pi} \int_{0}^{1}
    {dt\over\sqrt{t(1-t)}\sqrt{1+{p\over   p_{B}}t}}
    \end{equation}
which it is convenient to rewrite, to study the small $p_{B}$ limit, as 
\begin{equation}
G(p/p_{B})={1\over\pi}\sqrt{p_{B}/p}\int_{0}^{p/p_{B}}
{dt\over\sqrt{t(1+t)}\sqrt{1-p_{B}t/p}}
\end{equation}
From this we see that the leading corrections goes in fact as
$\sqrt{p_{B}/p}\ln(p_{B}/p)$ - in agreement with a perturbative
calculation near the UV fixed point.  The large $p_{B}$ limit can be studied as well, with 
\begin{equation} 
G(p/p_{B})=\sum_{k=0}^{\infty } {(-1)^{k}\over 2^{4k}}
{[(2k)!]^2\over (k!)^{4}} \left({p\over p_{B}}\right)^{k}
\end{equation}
such that $G=1$ when $p/p_{B}=0$. 
 The leading order is
linear in $p/p_{B}$, again in agreement with perturbative
calculations near the IR fixed point. It is  convenient to reexpress $G$ as a hypergeometric function
\begin{equation}
G(p/p_B)=F\left({1\over 2},{1\over 2},1;-{p\over p_B}\right).
\end{equation}
Standard formulas then allow for a complete expansion in the large $p_B$ limit. For further use, we quote the general result
\begin{eqnarray}
{\Gamma(1/2)\over \Gamma(c)}F\left({1\over 2},{1\over 2},c;-{p\over p_B}\right)={(p_B/p)^{1/2}\over \Gamma(c-1/2)} \sum_{n=0}^\infty
{\Gamma(1/2+n)\over\Gamma(1/2)}{\Gamma(3/2-c+n)\over\Gamma(3/2-c)}{1\over (n!)^2}\left(-{p\over p_B}\right)^{-n} \left[\ln(p/p_B)+h_n(c)\right]\label{bigguy}
\end{eqnarray}
where 
\begin{equation}
h_n(c)=2\psi(1+n)-\psi(n+1/2)-\psi(c-n-1/2)
\end{equation}

In the present case, $G$, after analytic continuation,
can directly be interpreted as the scattering amplitude for 
an incident electron into an outgoing electron, and $|G|^{2}$ is therefore 
the probability of this process
\begin{equation}
P_{1\rightarrow 1}=\left|G(ip/p_B)\right|^2=\left|F\left({1\over 2},{1\over 2},1;-i{p\over p_B}\right)\right|^2
\end{equation}
The result is also shown in figure \ref{figProb}.
    
Note that we can also extract from these results expressions for the 
density of states inside the Fermi liquid. This density has a
constant part and an oscillatory part, and thus reads, 
as a function of the distance $x$ from the point contact:
\begin{equation}
\rho(x)\propto \hbox{cst}+e^{ip_{F}x}\int~dp \ G(ip/p_{B})e^{ipx}+\hbox{h.c.}
\end{equation}
This will be discussed in more details elsewhere.

\subsection{The case $lqp\rightarrow lqp$}

When a Landau quasi-particle (lqp) scatters into a Landau quasi-particle,
the amplitude is related to a slightly different kind of
correlator, 
\begin{eqnarray}
G(4,2)&=&g(4,2)~\langle
T~\exp(i(\alpha/\sqrt{3})\phi'_L(-x_4,t_4))
\exp(-i(\alpha/\sqrt{3})\phi'_R(x_2,t_2))\rangle\nonumber\\
&=&g(4,2) ~{1\over
(x_4-x_{2}-t_{4}+t_2+i\epsilon~sign(t_{4}-t_{2}))^{\alpha^2/3}},
\end{eqnarray}
where $x_{4}>0,x_{2}<0$ and 
\begin{equation}
g(4,2)=\langle
T~\exp(-i\alpha\phi_L(-x_4,t_4))\exp(-i\alpha\phi_R(x_2,t_2))\rangle.
\end{equation}
The case $lqp\rightarrow lqp$ corresponds again to $\alpha=1/2$, that
is 
an expression which is identical, in the $\phi$ sector, to the
amplitude for the electron electron process. One can then deduce the 
lpq amplitude from the electron amplitude
\begin{equation}
    G(4,2)\propto \int_{0}^{\infty}p^{-2/3}e^{-px}dp 
    \int_{0}^{1} (1-u)^{-7/6}u^{-1/2}\left(1+{p\over
    p_{B}}u\right)^{-1/2}du\label{expiii}
    \end{equation}
where $x=x_{4}-x_{2}+i(y_{4}-y_{2})$. To obtain the latter result, we
used a Laplace representation of $x^{1/6}$, which is
divergent, and can be made finite by the analytical
continuation (continuation of a $\Gamma$ function in this case). 
Similarly, (\ref{expiii}) is divergent,  and can be made finite 
by expanding in powers of $p/p_{B}$ at low energy, and using $B$ (
$\Gamma$) functions regularization. One finds then
\begin{equation}
    G(p,p_{B})\propto p^{-2/3}\sum_{n=0}^{\infty}\left({p\over
    p_{B}}\right)^{n}(-1)^{n}{(2n)!\over 2^{2n}(n!)^{2}}
    {\Gamma(-1/6)\Gamma(n+1/2)\over \Gamma(n+1/3)}\label{deviii}
    \end{equation}
By applying Stirling's formula at large $n$, one can check that the series
converges for $p/p_B < 1$.
At large energy meanwhile, one finds the leading behaviour
\begin{equation}
    G(p,p_B)\propto p^{-2/3} \left({p_{B}\over p}\right)^{1/2}\ln(p_{B}/p)
    \end{equation}
again in agreement with perturbative calculations.  The complete large energy expansion can be obtained by recognizing that 
\begin{equation}
G(p,p_B)\propto p^{-2/3} F\left({1\over 2},{1\over 2},{1\over 3};-{p\over p_B}\right)
\end{equation}
and then using (\ref{bigguy}).

\begin{figure}
\vspace{0.3cm}
\begin{center}
\epsfig{file=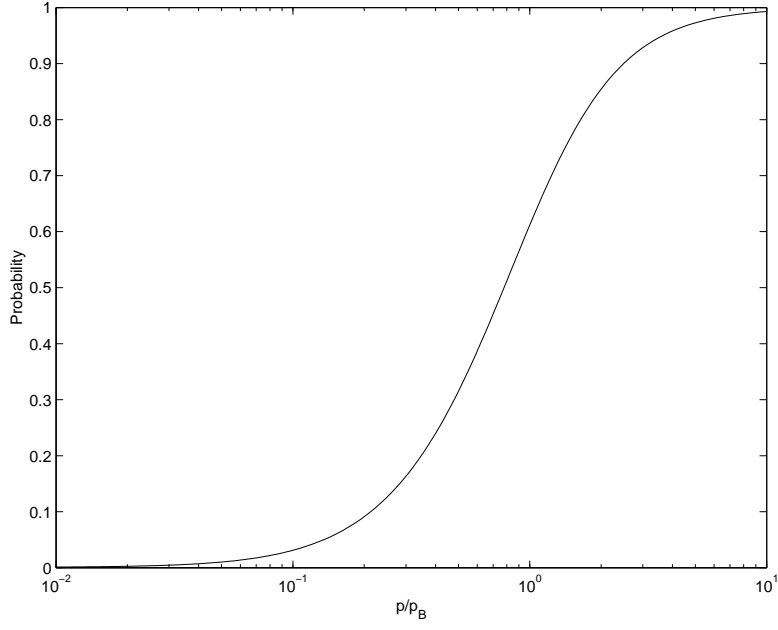,scale=0.6}
\caption{\label{figLpq} Probability for the process $lpq\rightarrow lpq$.}
\end{center}
\end{figure}

A graph of the associated probability (normalized by the low energy limit) i.e.
$|F({1\over 2},{1\over 2},{1\over 3};-{i p\over p_B})|^2$ is provided
in figure (\ref{figLpq}).
%

%
%

\subsection{The case $1e\rightarrow 1e_{edge}$.}

This is still another case, corresponding to $m=1,p=3$. The amplitude
is therefore
\begin{equation}
G(4,1)=g(4,1) ~{1\over
(x_4-x_{1}-t_{4}+t_1+i\epsilon~sign(t_{4}-t_{1}))^{3/4}},
\end{equation}
where $x_{4}>0,x_{1}<0$ and 
\begin{equation}
g(4,1)=\langle
T~\exp\left(-{3i\over 2}\phi_L(-x_4,t_4)\right)\exp\left({i\over 2}\phi_R(x_1,t_1)\right)\rangle.
\end{equation}
This amplitude is expected to vanish both at the UV and IR fixed
points, and take non trivial values in between. Note that the
amplitude is invariant under the shifts $\phi_{R}\rightarrow
\phi_{R}+\pi$ and $\phi_{L}\rightarrow \phi_{L}-\pi$ corresponding to
the UV boundary state, and $\phi_{R}\rightarrow\phi_{R}+2\pi$,
$\phi_{L}\rightarrow\phi_{L}+2\pi$ corresponding to the IR boundary
state. One can similarly check that the leading relevant (resp.
irrelevant) operator give it a non trivial value away from the UV
(resp. IR) fixed point. More sophisticated arguments presented in the
appendix give rise to the following expression
\begin{equation}
    G(4,1)\propto x^{-2}\int_{0}^\infty e^{-y}(1-2y) y^{-1/2}
    \left(1+{y\over p_{B}x}\right)^{-1/2} dy
    \end{equation}
from which the momentum dependent amplitude follows
\begin{equation}
    G(p,p_{B})\propto p^{1/4} \int_{0}^{1}
    (1-u)^{-1/4}u^{-1/2}{2-u\over 1-u}\left(1+{p\over p_{B}}u\right)^{-1/2}du
    \end{equation}
Note that this
correlator  vanishes in the low and high energy limits,
as expected from previous considerations about the fixed points.
An expansion at low energy  follows from $\Gamma$ function regularization
again
\begin{equation}
G(p,p_{B})\propto p^{1/4} \sum_{0}^{\infty}
(-1)^n \frac{(2n)!}{2^{2n}(n!)^2}
\left[ 2\frac{\Gamma (-1/4) \Gamma(1/2+n)}{\Gamma(1/4+n)}-
\frac{\Gamma(-1/4)\Gamma(3/2+n)}{\Gamma(5/4+n)}\right]
\left( \frac{p}{p_B}\right)^n
\end{equation}

The leading behaviour at high energy is
\begin{equation}
G(p,p_B)\propto p^{1/4}\left({p_B\over p}\right)^{1/2}\ln(p_B/p)
\end{equation}
Once again, one can recognize a sum of two gamma functions
\begin{equation}
G(p,p_{B})\propto p^{1/4}
\left[{\Gamma(1/2)\Gamma(3/4)\over \Gamma(5/4)} F\left({1\over 2},{1\over 2},{5\over 4},-{p\over p_B}\right)+{\Gamma(1/2)\Gamma(-1/4)\over\Gamma(1/4)}
F\left({1\over 2},{1\over 2},{1\over 4};-{p\over p_B}\right)\right]
\end{equation}
and then use formula (\ref{bigguy}) to obtain the complete high energy expansion. 

\begin{figure}
\vspace{0.3cm}
\begin{center}
\epsfig{file=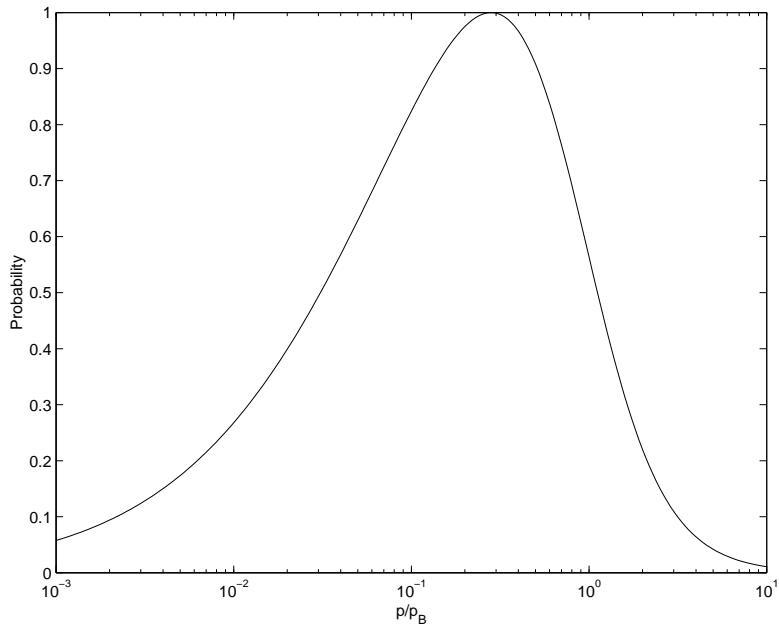,scale=0.6}
\caption{\label{figeedge} Probability (arbitrarily normalized) for the process $e\rightarrow e_{edge}$.}
\end{center}
\end{figure}

In the present case, we do not know how to normalize the associated probability, which is expected to vanish in both low and high energy limits - this would presumably require a more complete understanding of unitarity issues. Qualitatively, we expect that it should behave as $|G(ip,p_B)/p^{1/4}|^2$, which is  represented 
in figure (\ref{figeedge}).

\section{Conclusions}

The results we have obtained in closed form are but a very small step
towards a general understanding of the scattering amplitudes.
Extensions of this work to other problems would involve in general 
considerable technical
difficulties, if only because the form-factors approach is so hard to
implement in the massless case. On top of this, even the form-factors
themshelves might not be known, or only known in a very involved form
- clearly, a new idea is needed there if progress is to be made. 

A more rewarding direction of development would probably be to 
restrict to the present $g={1\over 2}$ case, and consider
 instead the question of
unitarity, and how the probability spreads among multiparticle
processes as energy is varied. It also remains to be seen whether a
direct determination of scattering amplitudes is experimentally
feasible.

\vskip2cm
\noindent{\bf Acknowledgments:} We thank C. de C. Chamon, 
E. Fradkin, N. Sandler and K. Schoutens for discussions. H.S. was supported by the Humboldt
foundation and the DOE. F. Lesage is supported by the NSERC.

\appendix
\section{Disorder operator and flow considerations.}

The one point function of the disorder operator in the Ising model
with a boundary magnetic field vanishes all along the flow. This
follows immediately from the fact that  all non vanishing
form-factors for $\mu$ involve odd numbers of particles, while the
boundary state is a superposition of an even number of particles only.

Here, a word of caution is necessary. Right at $h=0$, the proper boundary state in
the massive theory is in fact a superposition of states with odd and 
states with even numbers of particles:
\begin{equation}
    |B>_{free}=\left(1+Z^{\dagger}(\theta)\right)   
    \exp\left[{1\over 2}\int_{-\infty}^{\infty}K_{free}(\theta)
    Z^{\dagger}(-\theta)Z^{\dagger}(\theta)\right]
\end{equation}
One has $K_{free}=-i\coth{\theta\over 2}$, and the presence of
$Z^{\dagger}(0)$ is related with the pole of $K$ at $\theta=0$. So,
right at the free boundary conditions fixed point, the one point
function of the disorder operator is non zero. In the massless limit,
 the zero momentum particle becomes a zero
momentum zero energy particle, and becomes almost invisible except
for the fact that it changes odd into even number of particles, and
intorduces some  sign factors. One then finds (this is discussed in
more details below)
 \begin{equation}
     \langle\mu(x)\rangle_{h=0}=\sum_{n=0}^{\infty}{1\over n!}
     \int_{-\infty}^{\infty}\prod_{i=1}^{n}\left\{{d\beta_{i}\over
     2\pi}e^{-2xe^{\beta_{i}}}\right\}
     \prod_{i<j}\left(\tanh{\beta_{i}-\beta_{j}\over 2}\right)^{2}
     \end{equation}
 where the term $\prod_{i}\tanh{\beta_{B}-\beta_{i}\over 2}$ which 
 is there equal to $(-1)^{n}$, is cancelled by the $(-1)^{n}$ term
 arising from the presence of an additional particle at vanishing
 energy and momentum. This formal expression coincides with
 $\langle\sigma(x)\rangle_{h=\infty}$, and thus $\langle
 \mu(x)\rangle_{h=0}\propto x^{-1/8}$. 
   
For 
any non zero $h$ however, there is no one particle state in $|B>$
\cite{GZ}, and the one point function of $\mu$ vanishes exactly.  
 
In this appendix we first would like to discuss this result further. 
To do so, we first can try to 
 follow the route of \cite{CZ}, and derive a
 differential equation for $<\mu>$.  Recall that in the usual case, 
 Chatterjee and Zamolodchikov find the
 following equation 
 \begin{equation}
     \left[4{d^{2}\over dY^{2}}+\left({1\over Y}-4\right){d\over dY}+
     \left(-{1\over 2Y}+{9\over 16
     Y^{2}}\right)\right]\langle\sigma\rangle=0,~~Y=2p_{B} x
     \end{equation}
 ($x$ taken positive by convention here) and $p_{B}=4\pi h^{2}$.  Setting 
 $\langle\sigma\rangle=Y^{3/8}\Psi$, one finds the equation
 \begin{equation}
     Y\Psi''+(1-Y)\Psi'-{\Psi\over 2}=0
     \end{equation}
 whose physica lsolution is proportional to 
 the degenerate hypergeometric function
 $\Psi=\Psi(1/2,1,Y)$.

 In the  case of the disorder operator, the same logic as in \cite{CZ}
can be followed, with proper modifications due to the different
monodromies involves.  We find instead the equation
  \begin{equation}
      \left[4{d^{2}\over dY^{2}}+\left({1\over Y}-4\right){d\over dY}+
      \left(-{1\over 2Y}+{1\over 16
      Y^{2}}\right)\right]\langle\mu\rangle=0,~~Y=2 p_{B} x
      \end{equation}
 And, setting $\langle\mu\rangle=Y^{-1/8}\Psi$, one has this time
 \begin{equation}
     Y\Psi''-Y\Psi'=0
     \end{equation}
 with only solution (that does not grow exponentially at large
 distance)  $\Psi$ a constant. 
 The only physical solution is thus $<\mu>=0$, since
 we know it has to vanish in the limit of large fields, where the
 boundary spins are fixed (the case $p_{B}=0$ can be exceptional,
 since then the variable in the differential equation is not well defined).

 This calculation is formally the same as the calculation 
 of the one point function of the spin $\langle\sigma\rangle_{\tilde{h}}$ for an Ising model where the
 boundary perturbation, instead of being the boundary field, would
 induce the following equations of motion:
  (if $t$ is the
 coordinate along the boundary as in \cite{GZ})
 \begin{equation}
     \left({d\over dt}+ip_{B}\right)\psi(t)=-\left({d\over
     dt}-i p_{B}\right)\bar{\psi}(t)\label{schange}
     \end{equation}
Note the crucial minus sign on the right hand side - in the usual
problem of Ising model with a boundary magnetic field, the rhs comes 
with a plus sign, and the physics is the flow from free to fixed
boundary conditions. It is easy to check that switching the sign in
this equation leads to switching the sign of the reflection matrix,
which reads in this case 
\begin{equation}
    \tilde{R}=-i\tanh \left({\beta-\beta_{B}\over 2}-{i\pi\over
    4}\right)\label{newrmat}
    \end{equation}
    (in the following we set now $p_{B}=4\pi \tilde{h}^{2}$).  
    In the UV, one finds  $R=-i$ and in the IR $R=i$, so the role of the
 free and fixed boundary conditions are exchanged: this is the flow
 `dual' to the ordinary one, which can be
  understood for instance by using a low and high temperature graphical expansion
  of the Ising model partition functions. 
 
 Here, one has to be a bit careful.  From entropy considerations, one does not
  expect a flow from fixed to free boundary conditions to be
  possible. 
  Rather, what probably happens is 
  a flow from the superposition of two boundary states
  (fixed $+$ and fixed $-$) towards free boundary conditions. The
  ratio of degeneracies is then $g_{UV}/g_{IR}=2\times {1\over
  \sqrt{2}}=\sqrt{2}$, the same as for the usual flow from free to
  fixed. In such a flow, the $Z_{2}$ symmetry is never broken, and
  therefore $<\sigma>_{\tilde{h}}=0$ always. The free energy in this flow 
  is in fact the same (as a function of $T,p_{B}$) as the free energy 
  in the ordinary flow - a feature compatible with the fact that the
  free energy in the integrable approach only depends on the
  derivative of the log of the $R$ matrix.  In this point of view the 
  UV fixed point appears  extremely singular, as it is a
  superposition of two independent states - for any $p_{B}\neq 0$,
  these two states are coupled, and ``cutting the system in ($Z_{2}$) 
  half'' does not make physical sense. Only at
  the ``double fixed'' fixed point can one do so,
 and  define $\langle\sigma\rangle_{\tilde{h}=0}\propto x^{-1/8}$.

Now we can ask about the boundary state associated with
(\ref{newrmat}). In the usual case, the proper way to discuss this is
to go to the massive theory. This can be done here in various ways.
An interesting one is to make a detour through the massive Kondo
model \cite{Leclair}, that is the theory of massive fermions in the bulk, with a
Kondo like interaction at the boundary. This theory decouples into
two independent Ising models. One of them is the usual Ising model
with a boundary magnetic field, while the other has the following $R$
matrix
\begin{equation}
    \tilde{R}=-i\coth\left({i\pi\over 4}-{\theta\over 2}\right){1+{\tilde{h}^2\over
    2m}+i\sinh\theta\over 1+{\tilde{h}^{2}\over 2m}-i\sinh\theta}
    \end{equation}
Here one sees that $\tilde{R}=R_{free}$ at large value of the coupling $h$,
while $\tilde{R}=R_{fixed}$ at small value of the coupling. For any value of 
the coupling, the $K$ matrix has a pole at the origin,
indicating the presence of a zero momentum particle in the boundary
state. 
    Note that in the massless limit the two reflection matrices become  
    $R=i\tanh\left(
{\beta-\beta_{B}\over 2}-{i\pi\over 4}\right)$, and  $\tilde{R}=-i\tanh\left(
{\beta-\beta_{B}\over 2}-{i\pi\over 4}\right)$. The model thus
decomposes into two Ising models, one with a flow from free to fixed,
and one with a flow from ``double fixed'' to free (note that the ratio of $g$
factors is thus $g_{UV}/g_{IR}=\sqrt{2}\times {2\over \sqrt{2}}=2$ as
required for the spin $1/2$ Kondo model.)  The boundary state for the 
latter flow always contain a zero momentum, zero energy particle

%
    %

To proceed, let us
consider the form factor 
\begin{equation}
   \left<0|\mu|0,\theta_{2n},\ldots,\theta_{1}\right>=
 \prod_{i=1}^{2n}\tanh(\theta_{i}/2)
   \prod_{i<j=1}^{2n}\tanh(\theta_{ij}/2)
   \end{equation}
 In the massless limit, $n$ particles will become R movers, $n$
 will become L movers. Whatever particles we choose for this, the
 particle at rapidity zero will introduce a factor $(-1)^{n}$, and
 multiplying this factor, we will have the massless limit of the
 product $\prod_{i<j=1}^{2n}\tanh(\theta_{ij}/2)$, which will  be a factor of similar terms for R and L components
 independently, times a sign factor which depends on which particles 
 become R and which particles become L. The minus sign hence
 generated by the particle at vanishing rapidity exactly cancels the 
 minus sign coming from the $\tilde{R}$ matrix, so we can write at the end
 \begin{equation}
     \langle\mu\rangle_{\tilde{h}}=\langle\sigma\rangle_{h}
     \end{equation}
hence establishing that the one point function of the disorder
operator in the dual flow is the same as the one point function of
the order operator in the original flow.

  A last comment about the dual flow. 
  Although we do not know the action corresponding to the one point
function near the fixed boundary conditions fixed point 
(and thus the parameter $\tilde{h}$ is only formal), we can find
it near the IR (free) boundary conditions fixed point by using the
arguments of \cite{LS}. Since the $\tilde{R}$ matrix is analytic,
it follows that this action involves only terms of the form
$\psi\partial^{2n-1}\psi$. Since on the other hand the correlation
functions of the spin operator with these (conserved) quantities all 
vanish with free boundary conditions, it follows that, in a
perturbative expansion near the free boundary conditions (IR) fixed
point, $\langle\sigma\rangle_{\tilde{h}}$ has to vanish identically.
Of course, it may well be that the expansion of this quantity 
has a vanishing radius of convergence, wiht  non perturbative
contributions - the foregoing arguments have established that this is
not the case however.


As discussed in the main text, a regularized version of $\langle\mu\rangle_{h}$
can be obtained by adding to the boundary state a fermion at
vanishing energy and momentum. The corresponding expression in terms 
of form-factors is given in (\ref{howpuzzling}). 

An exact expression for this can be obtained by considering, 
following Chatterjee Zamolodchikov \cite{CZ}, the correlator,
 $<\chi(z)\mu(w,\wbar)>$
where $\chi\equiv (\partial_z+i\lambda)\psi$ for $Im \ z > 0$ which we
assume in the following. To start, 
we consider in fact a simpler correlator, $<\psi(z)\mu(w,\wbar)>$. 
We calculate this correlator using the form factor formalism, after 
insertion of the boundary $|B>$. The $n^{th}$ order term then reads
$$
{1\over n!}\int_{-\infty}^\infty \prod_{i=1}^n d\beta_i (-i) \tanh{\beta_B-\beta_i\over 2}
(-1)^{n(n-1)/2}
<0|\psi(z)\mu(w,\bar{w})|\beta_1,\ldots,\beta_n;\beta_1,\ldots,\beta_n>_{R,\ldots,R,L,\ldots,L}
$$
where the $(-1)^{n(n-1)/2}$ originates from reordering the $n^{th}$
temr in the expansion of the boundary state. Note that this term is
exactly cancelled by the sign coming from the massless limit of the
$\mu$ form factor. 

The disorder operator has non vanishing form factors only on odd numbers of particles. 
The fermion is simply able to destroy one particle. 
Suppose $\mu$ destroy the particles $\beta_2,\ldots,\beta_n$ in the R channel and $\beta_1,\ldots,\beta_n$ in the R channel. 
Then $\psi$ the remaining particle $\beta_1$  in the R channel, so the matrix elements $<0|...>$ will give the contribution
$$
e^{\beta_1/2} i^n\prod_{i<j=2}^n \tanh{\beta_i-\beta_j\over 2}
\prod_{k<l=1}^n \tanh{\beta_k-\beta_l\over 2}
\exp\left[-2x\left(e^{\beta_2}+\ldots+e^{\beta_n}\right)\right]
\exp[-(z-w) e^{\beta_1}]
$$
and $x={w+\wbar\over 2}$. 

Let us now look at the large $z$ behaviour ($Im \ z > 0$) of the corresponding integral. In that limit,
the integral is dominated by the region where $\beta_1\rightarrow -\infty$, and we
are left with the integral over $\beta_2,\ldots\beta_n$ of the expression
$$
i^n\prod_{i<j=2}^n \tanh{\beta_i-\beta_j\over 2}
(-1)^n\prod_{k<l=2}^n \tanh{\beta_k-\beta_l\over 2}
\exp\left[-2x\left(e^{\beta_2}+\ldots+e^{\beta_n}\right)\right]
$$
while the integral over $\beta_1$ produces a factor $1/\sqrt{z}$. 
The crucial factor $(-1)^n$ comes from the limit of the form factors contributions $\tanh{\beta_1-\beta_i\over 2}$. 
The factor coming from the boundary state meanwhile reduces to $\prod_{i=2}^n
\tanh{\beta_B-\beta_i\over 2}$. 

Now we could also have $\psi$ annihilate any of the other $\beta_2,\ldots\beta_n$ R particles instead. 
If this so happens, commuting $\psi$ through the particles on the left of the ones it annihilates generates minus signs. 
These however will be cancelled exactly by the minus signs coming from the limit of the product of $\tanh$ factors. 
It thus simply follows that we get a multiplicity of $n$ for the term we have just analyzed, and therefore at large $z$ 
\begin{eqnarray}
<\chi(z)\mu(w,\wbar)>\propto \lambda\sum_{n=0}^\infty
\int_{-\infty}^\infty 
\prod_{i=1}^n {d\beta_i\over 2\pi} \left(\tanh{\beta_B-\beta_i\over 2}\right)^2\nonumber\\
(-1)^n
\prod_{i<j=1}^n \left(\tanh{\beta_i-\beta_j\over 2}\right)^2
\exp\left[-2x\left(e^{\beta_1}+\ldots+e^{\beta_n}\right)\right]\nonumber
\end{eqnarray}
and the proportionality factor is independent of $\lambda$. Of course 
as always the generic terms in the series are divergent, and we do not learn
much (in particular, the $z$ power law dependence is not reliable)
until
we regularize by taking ratios. 

Using \cite{CZ}, we can on the other hand calculate the
large $z$ behaviour by computing $B$ in the right hand side of their
 equation (24). Putting things together we obtain the final result that
\begin{eqnarray}
\left(\sum_{n=0}^\infty (-1)^{n}
\int_{-\infty}^\infty 
\prod_{i=1}^n {d\beta_i\over 2\pi}
(\tanh{\beta_B-\beta_i\over 2})^2\prod_{i<j=1}^n \left(\tanh{\beta_i-\beta_j\over 2}\right)^2
\exp\left[-2x\left(e^{\beta_1}+\ldots+e^{\beta_n}\right)\right]\right)\nonumber\\
/
\left(\hbox{same with $\beta_B=\infty$}\right)=
-\sqrt{z}\left[2{d\over
dz}\Psi(1/2,1;z)-\Psi(1/2,1;z)\right],~~z=2p_{B} x\label{ratioi}
\end{eqnarray}
Numerical checks of this result (obtained by carrying out the form
factors expansion to fourth order) are presented in the figure. The 
agreement is not spectacular, but maybe this is expected, as 
successive orders in the form-factors expansion all add up with the 
same (positive) sign, while in the usual case they form an
alternating series. Note the divergence at vanishing $p_{B}$.
\begin{figure}
\vspace{0.3cm}
\begin{center}
\epsfig{file=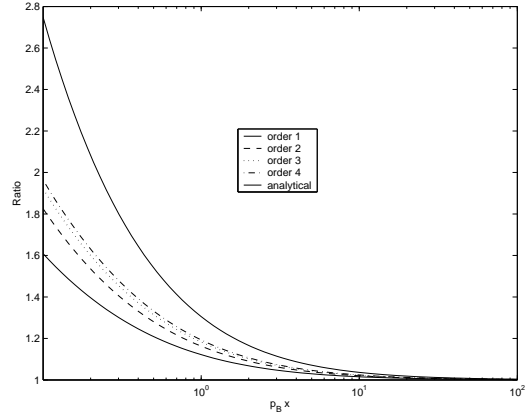,scale=0.4}  
\caption{\label{figappendix} Exact result and form factors estimates for the Ratio (\ref{ratioi})  as a function of $p_B$.}
\end{center}
\end{figure}

\section{Analytical relations.} 
 
This appendix details a simple extension of the method of 
Chatterjee-Zamolodchikov \cite{CZ}  to determine  correlators directly in the sine-Gordon
model. 
Let us start from
\begin{equation}
\label{Lagrangien}
A = \frac{1}{8\pi}\int_{-\infty}^0 dx \ [(\partial_x\Phi)^2-(\partial_t\Phi)^2]
+ v \cos \frac{1}{2} \Phi(x=0,t).
\end{equation}
Fermionizing 
using (we revert to the more standard notation $\psi,\bar{\psi}$)$\psi_\pm (z)=e^{\pm i\phi(z)}$ and  $\bar{\psi}_\pm (\bz)=e^{\mp i\bar{\phi}(\bz)}$.
we get an action consisting of two decoupled Ising models having an interaction at the boundary
(see \cite{LeClairFermions} for details).
Taking the variation with respect to the boundary action, we find 
boundary conditions for the fermionic 
fields 
\begin{eqnarray}
\psi_+(z) + \psi_-(z) & = & \bpsi_+(\bz)+\bpsi_-(\bz) \\
i\partial_t (\psi_+-\psi_-)+v (\psi_+-\psi_-) & = & 
 -i\partial_t (\bpsi_+-\bpsi_-)+v (\bpsi_+-\bpsi_-).
\end{eqnarray}
Following Chatterjee-Zamolodchikov, we define then 
\begin{eqnarray}
\chi_+(z) & = & \psi_+(z)+\psi_-(z)  \\
\chi_-(z) & = & \frac{i}{2}\partial_z (\psi_+-\psi_-)+v(\psi_+-\psi_-) 
\end{eqnarray}
with $z=\frac{1}{2}(t+ix)$.  The boundary conditions  
insure that the fields are analytic
in the full complex plane. In terms of the bosonic field they read
\begin{eqnarray}
\chi_+(z) & = & 2\cos\phi(z)  \\
\chi_-(z) & = &-\partial_z \sin\phi(z)+ i v \sin\phi(z).
\end{eqnarray}
Now, given the OPE of the operators $\chi_\pm$ with the vertex operators
$V_{1/2,-1/2}$
we expect that 
\begin{eqnarray}
\langle\chi_+(z)e^{i( \phi(w)-i\bphi(\bar{w}))/2}\rangle (z-w)^{1/2} (z-\bar{w})^{1/2}
 & = &  B_+(w,\bar{w}) \\
\langle\chi_-(z)e^{i( \phi(w)-i \bphi(\bar{w}))/2}\rangle (z-w)^{1/2} (z-\bar{w})^{1/2}
& = &  \frac{A_-}{(z-w)}+ \\ \nonumber
 \frac{\bar{A}_-}{(z-\bar{w})}&+&B_-(w,\bar{w})\label{fredana}
\end{eqnarray}
The first equation for instance follows from analyticity together
with the OPE around $z=w$ 
\begin{eqnarray}
 \chi_+(z) V_{1/2,-1/2}(w,\bar{w} )& \simeq & \frac{1}{(z-w)^{1/2} }
\left[  V_{-1/2,-1/2} \right.\\
 & +& \left. (z-w) (V_{3/2,-1/2} - i \partial\phi V_{-1/2,-1/2}+\cdots )\right]
\end{eqnarray}
 a similar  OPE around $z=\bar{w}$, and the condition
 $\chi_{+}\approx {1\over z}$ at infinity. 

Expanding the first equation in (\ref{fredana}) in powers of $z-w$ leads to the
determination of  $B_+=(w-\bar{w})^{1/2} \langle
V_{-1/2,-1/2}(w,\bar{w})\rangle$.
The comparison with the expansion in $(z-\bar{w})$ leads to a tautology
relating the one point function of $V_{1/2,1/2}$ and that of
$V_{-1/2,-1/2}$.

Expanding to higher orders gives non trivial relations. 
For example, to next order in $(z-w)$ we have
\begin{equation}
\left\langle -\frac{B(w,\bar{w})}{2 (w-\bar{w})^{3/2}}  \right\rangle  =  
\langle V_{3/2,-1/2} \rangle - 
i \langle \partial\phi V_{-1/2,-1/2}\rangle  
\end{equation}
leading to the simple
\begin{equation}
\label{togoodtobetrue}
  \langle V_{3/2,-1/2} \rangle=-\frac{1}{2 x} \left\langle 
V_{-1/2,-1/2} \right\rangle
-2\frac{\partial}{\partial x}  \langle V_{-1/2,-1/2}\rangle
\end{equation}
giving an exact result for $\langle V_{3/2,-1/2} \rangle$ in terms of
$\langle V_{-1/2,-1/2} \rangle$. Since the latter quantity is known
from \cite{CZ}, the former follows.

\end{document}